\documentclass[preprint,12pt]{elsarticle}

\usepackage[margin=1in]{geometry} 
\usepackage{natbib}
\usepackage{amsmath}
\usepackage{amsfonts}
\usepackage{amssymb}
\usepackage{graphicx}
\usepackage{graphicx}
\graphicspath{{./figures/}}  
\usepackage{subcaption}
\usepackage{multirow}
\usepackage{soul}
\usepackage{color}
\usepackage[normalem]{ulem}
\usepackage[colorlinks=true,linkcolor=blue,citecolor=blue
]{hyperref}
\usepackage{algorithm}
\usepackage{algpseudocode}
\usepackage{booktabs}
\usepackage{array} 
\usepackage[T1]{fontenc}
\usepackage{placeins}
\newcolumntype{C}[1]{>{\centering\arraybackslash}m{#1}}
\newcolumntype{P}[1]{>{\centering\arraybackslash}p{#1}}
\usepackage{scalerel}

\begin{document}
\begin{frontmatter}

    \title{Physics-constrained Gaussian Processes for Predicting Shockwave Hugoniot Curves}
    
    \author{George D. Pasparakis\corref{cor1}\fnref{label1}}
    \author{Himanshu Sharma\fnref{label1}}
    \author{Rushik Desai\fnref{label2}}
    \author{Chunyu Li\fnref{label2}}
    \author{Alejandro Strachan\fnref{label2}}
    \author{Lori Graham-Brady\fnref{label1}}
    \author{Michael D. Shields\fnref{label1}}
    
    \address[label1]{Department of Civil and Systems Engineering, Johns Hopkins University, Baltimore, Maryland, USA}
    \address[label2]{School of Materials Engineering, Purdue University, West Lafayette, Indiana, USA}
    \cortext[cor1]{Corresponding author}

\begin{abstract}
A physics-constrained Gaussian Process regression framework is developed for predicting shocked material states and their associated uncertainties along the Hugoniot curve using data from a small number of shockwave simulations. The proposed Gaussian process is constrained by the Rankine–Hugoniot jump conditions between the various shocked material states to construct a thermodynamically consistent covariance function. This leads to the formulation of an optimization problem over a small number of interpretable hyperparameters and enables the identification of regime transitions, from a leading elastic wave to trailing plastic and phase transformation waves. Shock Hugoniots are an important measure for understanding material behavior under extreme conditions, including for the development of equations of state and determining material properties such as the Hugoniot Elastic Limit, but they are costly to generate through 
large-scale molecular dynamics simulations or shock experiments. 
Under these constraints, the proposed methodology establishes Hugoniot curves from a limited number of molecular dynamics simulations. We consider silicon carbide as a representative material and Molecular Dynamics simulations are performed using a reverse ballistic approach. The framework reproduces the Hugoniot curve with satisfactory accuracy while also quantifying the uncertainty in the predictions using the Gaussian Process posterior. These uncertain Hugoniot predictions can then be used to calibrate equation of state models, estimate material properties, or inform future experimental and/or simulation campaigns.
\end{abstract}
\end{frontmatter}


\section{Introduction}
The measured shock response of a material provides critical insight into 
a broad range of mechanical and thermodynamic phenomena, including elasticity, plasticity, phase transformations, and distinct failure mechanisms, while revealing critical features of the underlying equation of state.  Both experimental and computational methods can be used to probe the shock response of the material. 
Experimental characterization of shock response poses numerous challenges, such as sensitivity to loading parameters such as sample confinement, impact angle, sample alignment, and crystal orientation relative to the shock, resulting in a large number of experimental scenarios~\cite{vogler2006hugoniot}; not to mention the large expense and challenges/errors in measurements of (essentially) discontinuous states at extremely fast time scales. 

Computational methods, and Molecular dynamics (MD) specifically, 
can provide a detailed window into the underlying physics of the shock response~\cite{hamilton2021chemistry}. Early work on metals revealed the atomistic mechanisms behind plastic deformation \cite{holian1998plasticity}, phase transitions \cite{kadau2002microscopic} and dynamic failure \cite{strachan2001critical} under shock conditions. 
Despite demonstrating good agreement with experimental data~\cite{kadau2007shock}, MD simulations can be extremely computationally demanding, making them impractical for large-scale investigations or studies that require numerous simulations. 

The large expense and challenges associated with both experiments and MD simulations limit the insights we can gain into the thermodynamic response of materials in extreme conditions. There is a critical need to build basic thermodynamic models that generalize across shock conditions, satisfy essential thermodynamic laws, and can be learned from limited experimental and simulation data. Machine learning methods offer great promise in satisfying this need. 
For example, neural network-based approaches have been developed for modeling interatomic potentials \cite{behler2016perspective}, generating samples from equilibrium distributions \cite{noe2019boltzmann}, predicting full-field material responses \cite{pasparakis2025bayesian}, and mapping microstructure to thermal response under shock loading \cite{li2023mapping}. However, neural network-based approaches require a large amount of training data, which is often difficult or infeasible to obtain. Furthermore, these approaches lack essential uncertainty quantification (UQ) capabilities that provide a measure of prediction confidence, can be used to inform optimal design of experimental and simulation studies, and can be propagated to assess uncertainty in downstream tasks such as large-scale hydrocode simulations.

To address this critical challenge, we propose a thermodynamically constrained Gaussian process regression (GPR) framework as a data-efficient, non-parametric Bayesian model for the shock response of a material with an inbuilt measure of uncertainty.
This work builds on recent studies \cite{gaffney2022constraining,sharma2024learning} introducing physics-constrained GPR frameworks for data-driven materials modeling under extreme temperature and pressure conditions. 
Motivated by these developments, a physics-constrained GPR framework is proposed for predicting shocked material states along the Hugoniot. 
This approach applies to a broad class of materials by adhering to fundamental conservation laws across the shock front. Specifically, the framework enforces the Rankine–Hugoniot conditions as physical constraints by deriving a physics-based covariance function that inherently satisfies these conditions. The proposed method is capable of modeling the transition from purely elastic shock response to elastic response with a trailing plastic wave, and ultimately into the overdriven regime. Unlike existing empirical models with fixed functional forms~\cite{knudson2013adiabatic,ye2024shock}, the proposed model has a flexible GPR functional form that relies on a small number of interpretable hyperparameters and is analytically shown to be consistent with the laws of thermodynamics. This allows it to be used to predict new, unobserved material states with an inbuilt measure of uncertainty. As a result, it can be used to predict important material properties such as the Hugoniot Elastic Limit (HEL) and can be directly applied, for example, to develop equation of state models for broad classes of materials using methods such as those developed by Sharma et al.~\cite{sharma2024learning}.

The model is applied to Silicon Carbide (SiC) as a representative material. SiC is a highly versatile engineering material that possesses exceptional physical properties, such as high thermal conductivity, stiffness, and hardness along with low density that make it widely used for applications involving extreme conditions. It is shown that the proposed method can model SiC Hugoniot behavior in a computationally inexpensive manner using limited MD simulation data while also quantifying the uncertainty in the estimated state.

\section{Background}\label{problem_statement}
\begin{figure}[!ht] 
    \centering
    \begin{subfigure}{0.49\textwidth}
        \centering
        \includegraphics[height=1.9in]{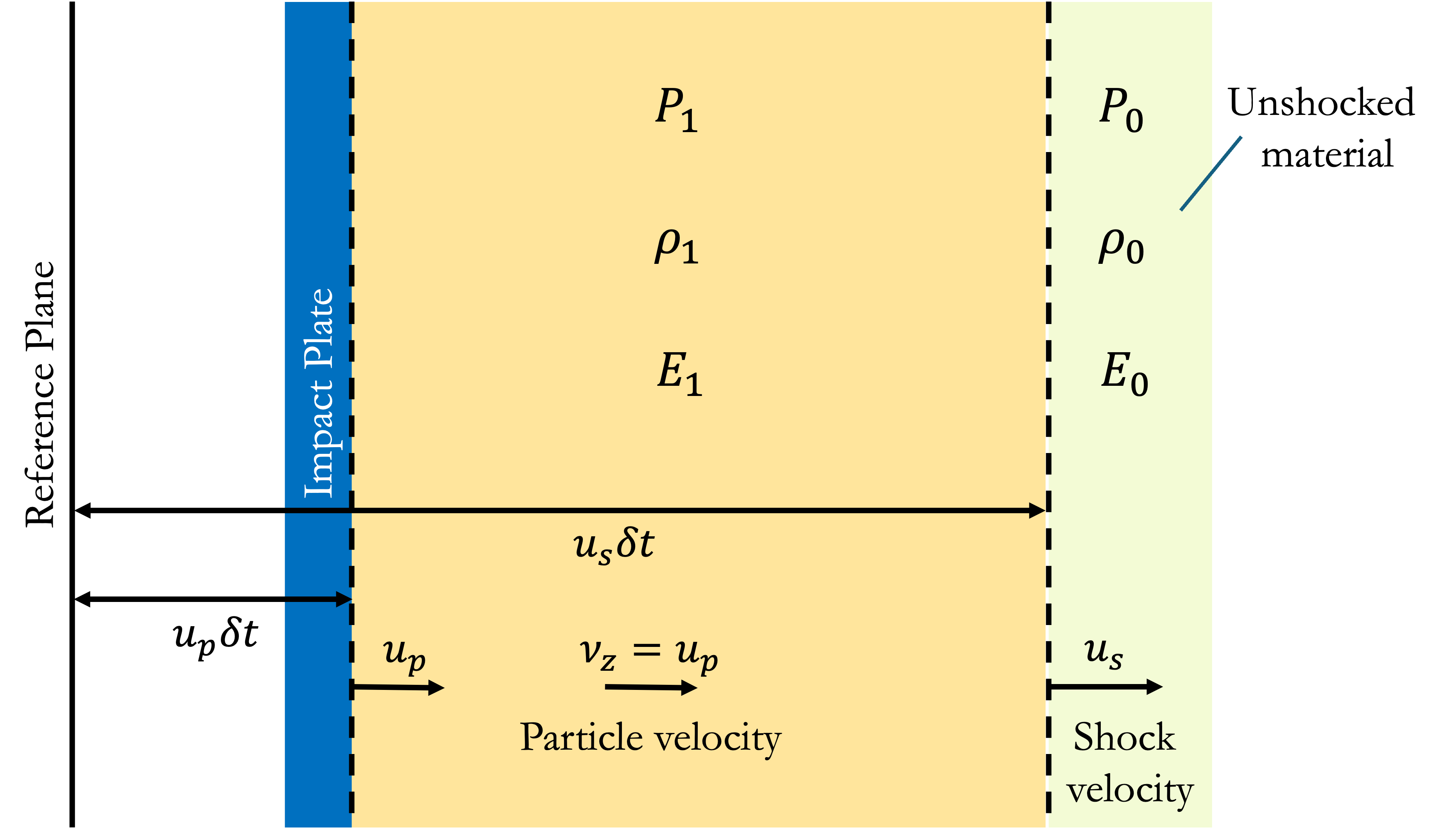}
        \caption{}
        \label{schematic_shock_1}
    \end{subfigure}
    \begin{subfigure}{0.49\textwidth}
        \centering
        \includegraphics[height=1.9in]{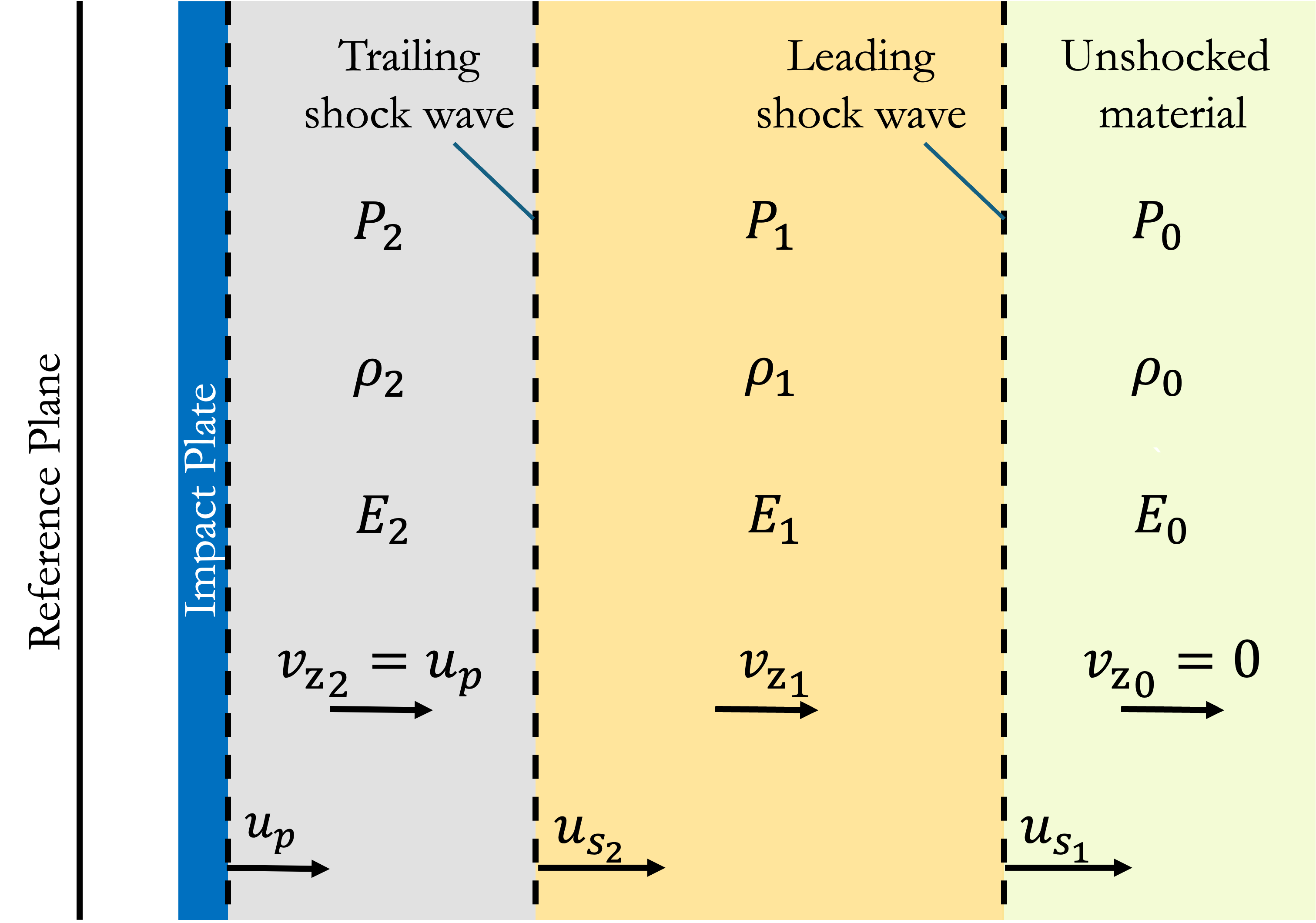}
        \caption{}
        \label{schematic_shock_2}
    \end{subfigure}
    \caption{(a) Schematic representation of shock wave propagation in a material element below the HEL (b) Schematic of shock wave propagation above the HEL resulting in a two-wave structure.}
    \label{schematic_shock}
\end{figure}

Shock-compression occurs when a material is subjected to an abrupt, high-speed impact or loading event, whose duration is considerably shorter than the characteristic time for the material to respond through inertial motion. For sufficiently high impact velocities, a structured wave with very short rise time (sub-picoseconds within single crystals) forms such that the thickness of the wave front can be ignored or cannot be measured experimentally. This leads to a jump in the principal mechanical and thermodynamic variables such as pressure, strain, velocity, and temperature. A schematic representation of shock wave loading is shown in Figure~\ref{schematic_shock_1}, where a projectile impacts the material at piston velocity $u_p$. The shock wave travels through the medium with velocity $u_s$, which is hypersonic relative to the unperturbed material, while the material behind the shock front moves at the particle velocity $\nu_z$ relative to a stationary reference frame (shown here as the initial boundary of the undeformed material).

For steady-state shocks, applying conservation of mass, momentum, and energy across the jump yields
the Rankine-Hugoniot jump conditions, in which the shock wave is characterized by a jump in pressure, density, and temperature across the wave front. For a single wave propagating in an unperturbed material, the Rankine-Hugoniot equations are given by \cite{asay2012high}:
\begin{equation}\label{hugoniot_pressure}
    P = \rho_0 u_s u_p + P_0
\end{equation}
\begin{equation}\label{hugoniot_density}
    \rho = \dfrac{\rho_0 u_s}{u_s - u_p}
\end{equation}
\begin{equation}\label{hugoniot_energy}
    E = E_0 + \dfrac{1}{2} \left( \rho_0^{-1} - \rho^{-1} \right) \left( P + P_0 \right)
\end{equation}
such that pressure \(P\), density \(\rho\), and internal energy \(E\) in the shocked material can be calculated from the initial unperturbed states  \(P_0\), \(\rho_0\), \(E_0\) together with the shock velocity \(u_s\) and the particle velocity \(u_p\). From a known ambient initial state, these three coupled equations can be solved for the shocked state by measuring \(u_s\) and \(u_p\) experimentally or from simulations.

A complete description of the thermodynamic state (i.e., closure of the system of equations) is achieved by introducing a fourth relation, typically in the form of an equation of state (EOS) model that relates the thermodynamic variables $P, V$ (or $\rho$), $T, E$ under shock conditions. Depending on the initial conditions, these relations define a set of thermodynamically accessible states in the \((P,V,T)\) space under shock conditions. Importantly, the temperature ($T$) in the shocked state cannot be derived directly from the Hugoniot equations without assuming an equation of state and measuring it experimentally is very difficult. However, temperature can be obtained from MD simulations or estimated from the relation \cite{hamilton2021chemistry}: 
\begin{equation}\label{energy_temperature_analytical}
    E - E_0 = \int_{T_0}^{T} C_v(T') \mathrm{d}T'
\end{equation}
where \(C_v(T)\) is the specific heat (which is assumed to be independent of pressure).

\begin{figure}[ht!]
    \centering
    \includegraphics[width=0.8\textwidth]{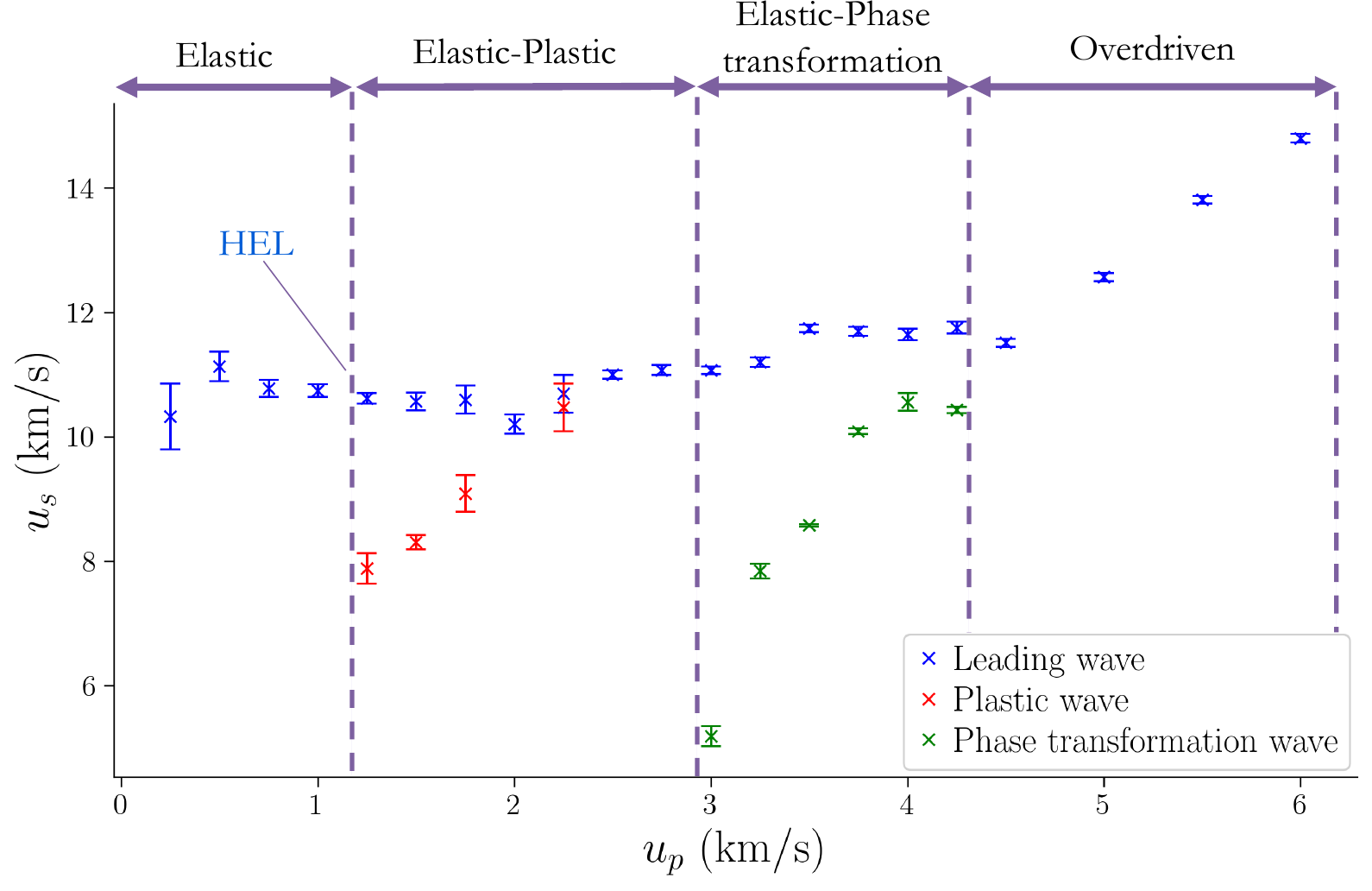}
    \caption{Shock velocity ($u_s$) vs. particle velocity ($u_p$) for single crystal SiC in the [0, 0, 1] orientation. Each `x' denotes a mean value +/- one standard deviation from MD simulations. Four regions are denoted. In the `Elastic' regime, a single elastic wave forms as shown by the blue x's. In the `Elastic-Plastic' regime, a leading elastic wave forms (blue x's) and is trailed by a plastic wave (red x's) whose shock velocity increases with increasing particle velocity until the two waves merge. In the `Plastic-Phase Transformation' regime, a leading plastic wave forms (blue x's) with a trailing phase transformation wave (green x's). In the `Overdriven' regime, a single phase transformation leading wave forms whose shock velocity increases with increasing particle velocity.}
    \label{Us_vs_Up}
\end{figure}

For relatively small particle velocities, a single shock wave propagates with velocity $u_s$ through elastic deformation in the material, as illustrated in Figure~\ref{schematic_shock_1} and shown in the left-most ``Elastic'' region in Figure~\ref{Us_vs_Up} relating $u_s$ and $u_p$ for SiC from MD shock simulations. When the imparted particle velocity induces stresses that result in plastic deformation, the material is said to reach its Hugoniot Elastic Limit (HEL). 
For imparted particle velocities above the HEL, a two-wave structure is often observed with an elastic leading wave (denoted by blue x's in the ``Elastic-Plastic'' regime in Figure~\ref{Us_vs_Up}) having shock velocity $u_{s_1}$ followed by a trailing plastic wave having shock velocity $u_{s_2}$ (denoted by red x's in Figure~\ref{Us_vs_Up}). Driving to higher particle velocities, $u_{s_2}$ increases and the trailing plastic wave and the elastic wave merge to form a single plastic leading wave (above $u_p\approx2$ km/s in Figure~\ref{Us_vs_Up}). At even higher particle velocities, additional thermodynamic processes are activated and
a trailing phase transformation wave forms (denoted by green x's in Figure~\ref{Us_vs_Up}), which again increases in velocity with higher particle velocity. Finally, in the ``Overdriven'' state, a single phase transition wave forms with increasing shock velocity. 

To account for the dual-wave structure that exists between elastic and overdrive shocks, illustrated in Figure~\ref{schematic_shock_2}, the Rankine-Hugoniot conditions are applied across each shock front as follows. 
Applying conservation of mass across the $i^{th}$ propagating wave front, we obtain 
\begin{equation}
    \rho_{i-1} (u_{s_i}-\nu_{z_{i-1}}) = \rho_{i} (u_{s_i}-\nu_{z_i})
\end{equation}
where $\nu_{z_i}$ represents the generalized particle velocity trailing the $i^{th}$ wave. Rearranging with respect to \(\rho_i\) yields
\begin{equation}
    \rho_{i} = \rho_{i-1} \frac{u_{s_i}-\nu_{z_{i-1}}}{u_{s_i}-\nu_{z_i}}
    \label{eqn:gen_hugoniot_density}
\end{equation}
Clearly, for the case of a single elastic wave where \(\nu_{z_0} = 0\) and $\nu_{z_1}=u_p$, Eq.~\eqref{eqn:gen_hugoniot_density} simplifies to
\begin{equation}\label{rho_gen}
    \rho_{1} = \rho_{0} \frac{u_{s_1}}{u_{s_1}-\nu_{z_1}} = \rho_{0} \frac{u_{s_1}}{u_{s_1}-u_p}
\end{equation}
which is identical to Eq.~\eqref{hugoniot_density}.
Likewise, applying conservation of momentum and energy to each shockwave yields the remaining generalized Hugoniot jump conditions given by
\begin{equation}
    P_i = \rho_{i-1}(u_{s_i} - \nu_{z_{i-1}})(\nu_{z_i} - \nu_{z_{i-1}}) + P_{i-1}
    \label{Hugoniot_pressure_gen}
\end{equation}
\begin{equation}
    E_i = E_{i-1} + \frac{1}{2}(P_i + P_{i-1})\left(\frac{1}{\rho_{i-1}} - \frac{1}{\rho_i}\right)
    \label{Hugoniot_energy_gen}
\end{equation}
Taken together, Eqs.~\eqref{eqn:gen_hugoniot_density}, \eqref{Hugoniot_pressure_gen}, and \eqref{Hugoniot_energy_gen} comprise the generalized Rankine-Hugoniot equations for multi-shock wave states. Note these generalized conditions may be applied to materials with any number of parallel shock fronts. For example, one might conceivably apply them to conditions where multiple successive shocks are imparted on the material. In this work, we will only consider the case where a single shock is imparted on the material with up to two shock fronts forming.

\section{Physics-constrained Gaussian Process for Shock State Prediction}\label{GP_formulation}

The objective of this work is to derive an uncertainty-aware machine learning model capable of predicting shock velocity, pressure, and other thermodynamic variables for a given shock scenario (i.e.\ a prescribed $u_p$) that is constrained by shock physics (i.e., the Rankine-Hugoniot equations in Eqs.~\eqref{eqn:gen_hugoniot_density}, \eqref{Hugoniot_pressure_gen}, and \eqref{Hugoniot_energy_gen}) and trainable from a limited number of MD simulations such as the data shown in Figure~\ref{Us_vs_Up}. This model will have broad applications supporting shock experiment efforts to predict shock states during experiment planning, predicting critical material properties such as the HEL, and providing thermodynamic state data for EOS development~\cite{sharma2024learning}. This machine learned shock model is derived herein using constrained Gaussian process regression, as described next.


\subsection{Gaussian Process regression}


A Gaussian process (GP), $Y(\mathbf{x})$, is a collection of random variables indexed on the input vector $\mathbf{x}$ such that, for any pair of inputs $\mathbf{x}$ and $\mathbf{x}'$, the pair $(Y(\mathbf{x}), Y(\mathbf{x}'))$ follows a joint normal distribution. 
As a result, a GP can be completely defined by its mean function  
\(\mu(\mathbf{x})\) and positive‑definite covariance function  
\(K(\mathbf{x},\mathbf{x}')\).
Given a set of training data \(\bigl\{\mathbf{x}^{(i)},y^{(i)}\bigr\}_{i=1}^{N}\) with \(\mathbf{X}=[\mathbf{x}^{(1)},\dots,\mathbf{x}^{(N)}]^{\top}\) and  \(\mathbf{y}=[y^{(1)},\dots,y^{(N)}]^{\top}\) the GP prior is written as
\begin{equation}\label{gp_prior}
  Y(\mathbf{x}) \sim
  \mathcal{GP}\bigl(\mu(\mathbf{x}),K(\mathbf{x},\mathbf{x}')\bigr)
\end{equation}
where \(\mu(\mathbf{x})\) is typically defined using a polynomial regression model.
A popular choice for the covariance function is the squared exponential (radial basis function) kernel with additive white noise given by,
\begin{equation}\label{gp_kernel}
  K(\mathbf{x},\mathbf{x}') =
  \sigma^{2}\exp\!\Bigl(-\tfrac{\|\mathbf{x}-\mathbf{x}'\|^{2}}{2\ell^{2}}\Bigr)
  \;+\; \sigma_{n}^{2}\delta_{\mathbf{x},\mathbf{x}'} 
\end{equation}
where \(\ell\) is the length‑scale,
\(\sigma^{2}\) the signal variance, and
\(\sigma_{n}^{2}\) the noise variance.
The hyper‑parameters are collected in the vector
\(\boldsymbol{\theta}=\{\ell,\sigma,\sigma_{n}\}\).

Let the kernel matrix be
\(\textbf{K} = [K(\mathbf{x}^{i},\mathbf{x}^{j})]_{i,j=1}^{N}\)  
and define the kernel entry
\(k(\mathbf{x}^{*}) = [K(\mathbf{x}^{(1)},\mathbf{x}^{*}),\dots,
                        K(\mathbf{x}^{(N)},\mathbf{x}^{*})]^{\top}\).
For a new input \(\mathbf{x}^{*}\) the posterior prediction conditioned on the training data $\mathbf{X, y}$ is Gaussian, and denoted by~\cite{williams2006gaussian} 
\begin{equation}\label{gp_posterior}
  Y(\mathbf{x}^{*}) \big| \mathbf{y},\mathbf{X},\mathbf{x}^{*}
  \;\sim\;
  \mathcal{N}\!\bigl(m(\mathbf{x}^{*}),s^{2}(\mathbf{x}^{*})\bigr)
\end{equation}
with posterior mean
\begin{equation}\label{gp_posterior_mean}
  m(\mathbf{x}^{*}) =
  \mu(\mathbf{x}^{*}) +
  k(\mathbf{x}^{*})^{\top} K^{-1}(\mathbf{y}-\boldsymbol{\mu})
\end{equation}
and posterior variance
\begin{equation}\label{gp_posterior_variance}
  s^{2}(\mathbf{x}^{*}) =
  K(\mathbf{x}^{*},\mathbf{x}^{*}) -
  k(\mathbf{x}^{*})^{\top} K^{-1} k(\mathbf{x}^{*})
\end{equation}
where
\(
  \boldsymbol{\mu}=[\mu(\mathbf{x}^{(1)}),\dots,\mu(\mathbf{x}^{(N)})]^{\top}.
\) 

The hyper‑parameters, $\boldsymbol{\theta}$, are usually inferred via a maximum‑a‑posteriori (MAP) estimate.
With a Gaussian prior
\(
  \boldsymbol{\theta} \sim \mathcal{N}(\boldsymbol{\mu}_{\boldsymbol{\theta}},\Sigma_{\boldsymbol{\theta}}),
\)
the MAP estimate is given by
\begin{equation}\label{MAP_objective}
  \hat{\boldsymbol{\theta}}_{\mathrm{MAP}} =
  \arg\min_{\boldsymbol{\theta}}\Bigl[
    -\log p(\mathbf{y}\mid\mathbf{X},\boldsymbol{\theta})
    \;-\; \log p(\boldsymbol{\theta})
  \Bigr]
\end{equation}
where the negative log‑likelihood term can be expressed as
\begin{equation}\label{NLL}
  -\log p(\mathbf{y}\mid\mathbf{X},\boldsymbol{\theta}) =
  \tfrac12 \Bigl[
    (\mathbf{y}-\boldsymbol{\mu})^{\top}
    K^{-1}(\mathbf{y}-\boldsymbol{\mu})
    + \log\lvert K\rvert
    + N\log(2\pi)
  \Bigr]
\end{equation}
and the negative log‑prior can be written as
\begin{equation}\label{prior_neglog}
  -\log p(\boldsymbol{\theta}) =
  \tfrac12 \Bigl[
    (\boldsymbol{\theta}-\boldsymbol{\mu}_{\boldsymbol{\theta}})^{\top}
    \Sigma_{\boldsymbol{\theta}}^{-1}(\boldsymbol{\theta}-\boldsymbol{\mu}_{\boldsymbol{\theta}})
    + \log\lvert\Sigma_{\boldsymbol{\theta}}\rvert
    + d\log(2\pi)
  \Bigr]
\end{equation}


\subsection{Rankine-Hugoniot constrained Gaussian process formulation} \label{Hugoniot_GP}\label{generalized_GP_model}
In recent years, it has become increasingly common to constrain GPR models to satisfy essential physical conditions~\cite{Swiler2020}.
In this section, we derive a constrained GP to predict the thermodynamic state variables \((P,V,T)\) and the shock velocity for a given piston velocity in a manner that is fully consistent with the the generalized Rankine-Hugoniot conditions in Eqs.~\eqref{eqn:gen_hugoniot_density}, \eqref{Hugoniot_pressure_gen} and \eqref{Hugoniot_energy_gen}.

First, a GP prior is placed on the two dependent variables \(u_{s_i}(u_p)\)  
and \(\nu_{z_i}(u_p)\) as 
\begin{equation}
    \boldsymbol{X}=\{u_{s_i}; \nu_{z_i}\}
    \;\sim\;
    \mathcal{GP}\Big(\boldsymbol{\mu}(u_p), \boldsymbol{K}(u_p, u_p')\Big)
    \label{eqn:X_def}
\end{equation}
where \(\boldsymbol{\mu}(u_p)\) and \(\boldsymbol{K}(\cdot,\cdot)\) are the mean vector and covariance (kernel) function for the outputs \(\{u_{s_i},\nu_{z_i}\}\). 
The auto- and cross-covariances are defined using the squared-exponential kernel in Eq.~\eqref{gp_kernel} as  
\begin{equation}
    \begin{aligned}
    K_{u_s u_s}(u_p,u_p) &= \sigma_{u_s}^2 \, k(u_p,u_p) \;+\; \sigma_{n,u_s}^2\,\delta_{u_p,u_p} \\[4pt]
    K_{v_z v_z}(u_p,u_p) &= \sigma_{v_z}^2 \, k(u_p,u_p) \;+\; \sigma_{n,v_z}^2\,\delta_{u_p,u_p} \\[4pt]
    K_{u_s v_z}(u_p,u_p) &= \rho\,\sigma_{u_s}\sigma_{v_z} \, k(u_p,u_p)
    \end{aligned}
    \label{eq:icm_blocks}
\end{equation}
These covariance components can be compactly expressed through the intrinsic coregionalization model (ICM) \cite{bonilla2007multi}, where the joint covariance is written as  
\begin{equation}
    \boldsymbol{K}(u_p, u_p) = \mathbf{B} \otimes k(u_p, u_p) + \mathbf{D}
    \label{eqn:icm_form}
\end{equation}
where \(\otimes\) denotes the Kronecker product, \(\mathbf{D}\) is a diagonal matrix whose \((l,l)\)-th element is the observation noise variance \(\sigma_{n,l}^2\) for the \(l\)-th output, and \(\mathbf{B}\) is a positive semidefinite coregionalization matrix given by  
\begin{equation}
    \mathbf{B} =
    \begin{bmatrix}
        \sigma_{u_s}^2 & \rho\,\sigma_{u_s}\sigma_{v_z} \\
        \rho\,\sigma_{u_s}\sigma_{v_z} & \sigma_{v_z}^2
    \end{bmatrix}
    \label{eqn:coreg_matrix}
\end{equation}
where \(\sigma_{u_s}\) and \(\sigma_{v_z}\) are the marginal standard deviations of \(u_s\) and \(\nu_z\), respectively, and \(\rho\) is their correlation coefficient (\(-1<\rho<1\)). Both outputs share a common squared-exponential kernel  
\begin{equation}
    k(u_p, u_p) = \exp\!\left[-\frac{(u_p - u_p)^2}{2\ell^2}\right]
    \label{eqn:shared_kernel}
\end{equation}
with a single characteristic length scale \(\ell\). The resulting model depends on a set of four hyperparameters, collected in the vector \(\boldsymbol{\theta} = \{\sigma_{u_s}, \sigma_{v_z}, \rho, \ell\}\).
To ensure thermodynamic consistency with the generalized Rankine-Hugoniot conditions in Eqs.~\eqref{eqn:gen_hugoniot_density}, \eqref{Hugoniot_pressure_gen}-\eqref{Hugoniot_energy_gen} 
we need to consider the following set of augmented outputs, $\boldsymbol{Y}=\{u_{s_i}, \nu_{z_i}, P_i, \rho_i, T_i\}^T,$
for which we define a joint GP prior by
\begin{equation}
\label{prior_gen}
\boldsymbol{Y} \Big| u_p
\;\sim\;
\mathcal{GP}\!\Bigl(
   \boldsymbol{\mu},\;
   \boldsymbol{\Sigma}
\Bigr)
\end{equation}
where \(\boldsymbol{\mu}\) is the mean vector for each component of~\(\boldsymbol{Y}\), and \(\boldsymbol{\Sigma}\) is the block covariance matrix:
\begin{equation}
    \label{cov_gen}
    \boldsymbol{\Sigma}
    =
    \begin{pmatrix}
    K_{{u}_{s_i},{u}_{s_i}} 
    & K_{{u}_{s_i},{\nu}_{z_i}}
    & K_{{u}_{s_i},{P}_i}
    & K_{{u}_{s_i},{\rho}_i}
    & K_{{u}_{s_i},{T}_i}
    \\[6pt]
    K_{{\nu}_{z_i},{u}_{s_i}}
    & K_{{\nu}_{z_i},{\nu}_{z_i}}
    & K_{{\nu}_{z_i},{P}_i}
    & K_{{\nu}_{z_i},{\rho}_i}
    & K_{{\nu}_{z_i},{T}_i}
    \\[6pt]
    K_{{P}_i,{u}_{s_i}}
    & K_{{P}_i,{\nu}_{z_i}}
    & K_{{P}_i,{P}_i}
    & K_{{P}_i,{\rho}_i}
    & K_{{P}_i,{T}_i}
    \\[6pt]
    K_{{\rho}_i,{u}_{s_i}}
    & K_{{\rho}_i,{\nu}_{z_i}}
    & K_{{\rho}_i,{P}_i}
    & K_{{\rho}_i,{\rho}_i}
    & K_{{\rho}_i,{T}_i}
    \\[6pt]
    K_{{T}_i,{u}_{s_i}}
    & K_{{T}_i,{\nu}_{z_i}}
    & K_{{T}_i,{P}_i}
    & K_{{T}_i,{\rho}_i}
    & K_{{T}_i,{T}_i}
    \end{pmatrix}
\end{equation}
The sub-blocks of \(\boldsymbol{\Sigma}\) associated with the auto-covariances and cross-covariances of \({P_i} \), \( {\rho_i} \), and \({T}_i\) are computed from the generalized Rankine-Hugoniot relations in Eqs.~\eqref{eqn:gen_hugoniot_density},\eqref{Hugoniot_pressure_gen} and \eqref{Hugoniot_energy_gen} using a Taylor series expansion around the mean vector \(\bar{\boldsymbol{x}}=\{{\bar{u}}_{s_i}, {\bar{\nu}}_{z_i}\}\) (i.e.\ the Delta method~\cite{casella2024statistical}) and considering known values for the state variables ahead of the shock wave, i.e., \({\nu}_{z_{i-1}},{\rho}_{i-1},{P}_{i-1}, {T}_{i-1},{E}_{i-1}\).

Consider that there exist $N$ input-output data points consisting of $\boldsymbol{u_p} = \{u_p^j\}_{j=1}^N$ and $\boldsymbol{Y}$ having components \(\boldsymbol{y}^j=\{u_{s_i}^j, \nu_{z_i}^j, P_i^j, \rho_i^j, T_i^j\}\) such that the Kernel functions of the block covariance matrix in Eq.~\eqref{cov_gen} are expressed as $N\times N$ matrices $\mathbf{K}_{lm}$ with $l,m=1,\dots, 5$ indexing the components of $\boldsymbol{Y}$. For brevity, the Hugoniot functions in Eqs.~\eqref{eqn:gen_hugoniot_density},\eqref{Hugoniot_pressure_gen} and\eqref{Hugoniot_energy_gen}
are written generically as functions $y_l=f_l\Bigl({u}_{s_i},{\nu}_{z_i}; \nu_{z_{i-1}}\Bigr)$ where again $l$ indexes the terms of $\boldsymbol{y}$ such that, e.g.\ $y_3=P_i$. This is justified in \ref{appendix_B} where it is shown that each of the Rankine-Hugoniot equations can indeed be written in this manner. Taking the Taylor series expansion of this abstracted function, the expectation operation \(\mathbb{E}[\cdot]\) is applied over the random variable, which yields the following mean value for the $j^{th}$ data point:
\begin{equation}
    \label{Taylor_mean_gen}
    \begin{aligned}
    \bar{y}_l^j
    \;=\;
    \mathbb{E}[y_l^j]
    &=
    f_l\Bigl(\bar{u}_{s_i}^j,\bar{\nu}_{z_i}^j; \nu_{z_{i-1}}^j\Bigr)
    \\[6pt]
    &\quad 
    + \frac{1}{2}
    \left[
    \left.\frac{\partial^2 f_l}{\partial (u_{s_i})^2}\right|_{\bar{\boldsymbol{x}}^j}
    \mathbf{K}_{u_{s_i}u_{s_i}}^{jj}
    +
    \left.\frac{\partial^2 f_l}{\partial (\nu_{z_i})^2}\right|_{\bar{\boldsymbol{x}}^j}
    \mathbf{K}_{\nu_{z_i}\nu_{z_i}}^{jj}
    \right.
    \\[4pt]
    &\qquad\quad \left.
    +
    2\left.\frac{\partial^2 f_l}{\partial u_{s_i} \partial \nu_{z_i}}\right|_{\bar{\boldsymbol{x}}^j}
    \mathbf{K}_{u_{s_i} \nu_{z_i}}^{jj}
    \right]
    \end{aligned}
\end{equation}
Likewise, the covariance between the \(j^{th}\) and \(k^{th}\) data points of the \(l^{th}\) and \(m^{th}\) output variables is given by 
\begin{equation}
    \begin{aligned}
        K\bigl(y_l^j, y_m^k\bigr) = \;&\; \mathbb{E}\bigl[y_l^j y_m^k\bigr]
    \;-\;
    \mathbb{E}[y_l^j] \mathbb{E}[y_m^k] \\[6pt] \approx\;&\;
        \left.\frac{\partial f_l}{\partial u_{s_i}}\right|_{\bar{\boldsymbol{x}}^j}
        \left.\frac{\partial f_m}{\partial u_{s_i}}\right|_{\bar{\boldsymbol{x}}^k}
        \mathbf{K}_{u_{s_i}u_{s_i}}^{jk} \\[6pt]
        &+
        \left(
        \left.\frac{\partial f_l}{\partial u_{s_i}}\right|_{\bar{\boldsymbol{x}}^j}
        \left.\frac{\partial f_m}{\partial \nu_{z_i}}\right|_{\bar{\boldsymbol{x}}^k}
        +
        \left.\frac{\partial f_l}{\partial \nu_{z_i}}\right|_{\bar{\boldsymbol{x}}^j}
        \left.\frac{\partial f_m}{\partial u_{s_i}}\right|_{\bar{\boldsymbol{x}}^k}
        \right) \mathbf{K}_{u_{s_i}\nu_{z_i}}^{jk} \\[6pt]
        &+
        \left.\frac{\partial f_l}{\partial \nu_{z_i}}\right|_{\bar{\boldsymbol{x}}^j}
        \left.\frac{\partial f_m}{\partial \nu_{z_i}}\right|_{\bar{\boldsymbol{x}}^k}
        \mathbf{K}_{\nu_{z_i}\nu_{z_i}}^{jk}
    \end{aligned}
    \label{Covariance_yjyk_final}
\end{equation}
The complete derivation for this covariance can be found in \ref{appendix_A}. Further, the final expressions for the mean values and each block term of $\boldsymbol{\Sigma}$ can be found in \ref{appendix_C} using the analytical derivatives from \ref{appendix_B}. 

Note that the augmented output vector $\boldsymbol{Y}$ contains temperature $T$ rather than internal energy $E$. As outlined in Section \ref{problem_statement}, temperature measurements are not typically available in shock experiments in which the specific heat \(C_v(T)\) is not readily available. To address this limitation, a linear temperature-energy relationship is assumed following the form 
\begin{equation}
    \label{energy_temp_poly}
    T_i = a + b E_i
\end{equation}
with covariance function expressed as
\begin{equation}
    \begin{aligned}
    K_{T_iT_i}^{jk}
    &\approx b^2K_{E_iE_i}^{jk}
    \end{aligned}
    \label{Covariance_TT}
\end{equation}
Likewise, the cross-covariances for temperature can be computed in a similar fashion as shown in \ref{appendix_C}. 

Each distinct sub-block of the full covariance matrix in Eq.~\eqref{cov_gen} can be expressed in compact form using the ICM. 
Specifically, using Eqs.~\eqref{eqn:coreg_matrix} and \eqref{eqn:shared_kernel}, the covariance term in Eq.~\eqref{Covariance_yjyk_final} for a pair of outputs \(y_l\) and \(y_m\) can be written as  
\begin{equation}
    \mathbf{K}_{lm}
    \;=\;
    \mathbf{L}_{lm}^\mathsf{T}
    \bigl(\mathbf{B} \otimes k(u_p,u_p)\bigr)
    \mathbf{L}_{lm}
    \;+\;
    \mathbf{D}_{lm}
    \label{eq:block_LBD}
\end{equation}
where \(\mathbf{L}_{lm}\) contains the first-order derivatives of the variables \(f_l(\cdot)\) and \(f_m(\cdot)\) with respect to the GP variables \(\{u_{s_i}, \nu_{z_i}\}\), as defined in Eq.~\eqref{Covariance_yjyk_final},  
\(\mathbf{B}\) is the coregionalization matrix from Eq.~\eqref{eqn:coreg_matrix}, and \(\mathbf{D}_{lm}\) is a diagonal matrix containing the observation noise variances for each output pair.  
The complete joint covariance matrix in Eq.~\eqref{cov_gen} can be compactly represented as the Kronecker product  
\begin{equation}
    \boldsymbol{\Sigma}
    \;=\;
    \mathbf{L}^\mathsf{T}
    \bigl(\mathbf{B} \otimes k(u_p,u_p)\bigr)
    \mathbf{L}
    \;+\;
    \mathbf{D}
    \label{eq:Sigma_tensor_form}
\end{equation}
where \(\mathbf{L}\) contains the first-order derivative terms from the Taylor expansion. Considering that both \(\mathbf{B}\) and \(\mathbf{K}_{u_p,u_p}\) are positive semidefinite, their Kronecker product is also positive semidefinite, and multiplication by \(\mathbf{L}^\mathsf{T}\) and \(\mathbf{L}\) preserves this property. Since \(\mathbf{D}\) is diagonal with nonnegative entries, it is positive semidefinite, and the sum of two positive semidefinite matrices remains positive semidefinite. This structured representation enforces joint positive semidefiniteness of \(\boldsymbol{\Sigma}\), thus improving numerical stability and mitigating matrix-invertibility issues that often arise in physics-constrained Gaussian Processes.

Furthermore, it can be ensured that the derived GP model satisfies the thermodynamic consistency and stability constraints expressed in ~\ref{app:Thermodynamic_Cosntraints}.
For the pressure stability constraint in Eq.~\eqref{stability_2}, the chain rule is applied as follows to show that, if the model satisfies the Rankine-Hugoniot equations (specifically Eq.~\eqref{Hugoniot_pressure_gen}), the stability constraint is automatically satisfied:
\begin{equation}\label{eq:stability1} 
    \dfrac{\partial P_i^j}{\partial V_i^j} = \dfrac{\partial P_i^j}{\partial \nu_{z_i}^j} \cdot \dfrac{\partial \nu_{z_i}^j}{\partial V_i^j} = \dfrac{\partial P_i^j / \partial \nu_{z_i}^j}{\partial V_i^j / \partial \nu_{z_i}^j} = -\bigl(\rho_i^{j}\bigr)^{2}\bigl(u_{s_i}^{j}-\nu_{z_i}^{j}\bigr)^{2} < 0
\end{equation}
Next, considering the linear model for temperature in Eq.~\eqref{energy_temp_poly}, differentiation with respect to \(E_i^j\) yields
\begin{equation}
    \frac{\partial T_i^j}{\partial E_i^j}=b
\end{equation}
and the consistency constraint in Eq.~\eqref{stability_1} is expressed as
\begin{equation}
    \frac{\partial E_i^j}{\partial T_i^j}=\frac{1}{b}
\end{equation}
Therefore, to ensure thermodynamic stability according to Eq.~\eqref{stability_1}, the polynomial coefficients \((a, b)\) are obtained by solving the following constrained optimization problem:
\begin{equation}
    \min_{a, b} \sum_{k=1}^N \left( T_i^k - (a + b E_i^k) \right)^2, \quad \text{subject to } b > \varepsilon \;\; \forall k
\end{equation}

Next, the model is trained using the input and the MD-generated vector of augmented outputs \(\boldsymbol{Y}\) (see \ref{sec:MD_simulations} for full details) to obtain the hyperparameters \(\boldsymbol{\theta}\) of the GP model using Eq.~\eqref{MAP_objective}. From a practical perspective, a straightforward use of the cross-covariance from Eq.~\eqref{cov_gen} leads to numerical instabilities as a result of poor relative scaling. In fact, physics-constrained GP models with multiple-outputs are known to suffer from positive definitiveness issues \cite{constantinescu2013physics}. In this work, the values associated with \(u_s\) and \(T\), for example, differ by several orders of magnitude such that \(\max(u_s) / \min(T) \sim \mathcal{O}(10^{-4})\), leading to large dynamic ranges across the sub-block covariances in Eq.~\eqref{cov_gen}. Furthermore, the analytical derivatives, which can dominate the overall magnitude of Eq.~\eqref{Covariance_yjyk_final}, lead to large dynamic ranges within the sub-block covariance terms. For example, \(\min(\boldsymbol{K}_{{T}_i,{T}_i}) / \max(\boldsymbol{K}_{{T}_i,{T}_i}) \sim \mathcal{O}(10^{-8})\). These scaling discrepancies result in \(\boldsymbol{\Sigma}\) being ill-conditioned and non-invertible. 
These issues are mitigated using a linear transformation of the diagonal elements of the Hessian such that they are approximately equal to one \cite{bertsekas1997nonlinear}. 

To ensure numerical stability, the following coordinate transformation is employed
\begin{equation}\label{scaled_values}
    \tilde{P}_i = s_1 {P}_i, \quad
    \tilde{{\rho}}_i = s_1 {\rho}_i, \quad
    \tilde{{T}}_i = s_2 {T}_i \quad
\end{equation}
where \(s_1, s_2\) are inversely proportional to the maximum values of \(\{{P}_i,{\rho}_i\}\) and \({T}_i\), respectively. 
It is readily seen that the original structure of the Rankine-Hugoniot conditions is preserved when using the scaling in Eq.~\eqref{scaled_values}. For practical purposes, training is then performed to obtain the optimal hyperparameters \(\boldsymbol{\theta}\) using the scaled augmented output \(\tilde{\boldsymbol{{Y}}}=\{u_{s_i},\;{\nu}_{z_i},\;\tilde{ P}_i,\;\tilde{{\rho}}_i,\;\tilde{T}_i\}^{\top}\) using gradient-based algorithms \cite{byrd1995limited}, although for clarity of the formulation the GP model is presented in terms of unscaled quantities.

For a given set of testing points \(\boldsymbol{u_p^*} = \{ u_p^{j} \}_{j=1}^M\) the predictive distribution of the joint GP model for the values \(\boldsymbol{Y}^* = \{{u_{s_i}^*}, {\nu_{z_i}^*},{P_i^*}, {\rho_i^*},{T_i^*}\}^{\mathrm{T}}\) is calculated as a joint Gaussian distribution across the training and test data points given by 
\begin{equation}
    \left.
    \begin{aligned}
    & \boldsymbol{Y} \\
    & \boldsymbol{Y}^*
    \end{aligned}
    \;\middle\vert\;
    \boldsymbol{u_p}, \boldsymbol{u_p^*}
    \right.
    \sim 
    \mathcal{GP}\left(
    \begin{bmatrix}
    \boldsymbol{\mu}(\boldsymbol{u_p}) \\
    \boldsymbol{\mu}(\boldsymbol{u_p^*})
    \end{bmatrix},
    \begin{bmatrix}
    \boldsymbol{\Sigma}(\boldsymbol{u_p}, \boldsymbol{u_p}) & \boldsymbol{\Sigma}(\boldsymbol{u_p}, \boldsymbol{u_p^*}) \\
    \boldsymbol{\Sigma}(\boldsymbol{u_p^*}, \boldsymbol{u_p}) & \boldsymbol{\Sigma}(\boldsymbol{u_p^*}, \boldsymbol{u_p^*})
    \end{bmatrix}
    \right)
    \label{eqn:joint_predictive}
\end{equation}
Conditioning on the training data gives the posterior predictive distribution as
\begin{equation}
    \begin{aligned}
    \boldsymbol{Y}^* \mid \boldsymbol{u_p}, \boldsymbol{Y}, \boldsymbol{u_p^*}
    \sim \mathcal{GP} \bigg(
    &\;\boldsymbol{\mu}(\boldsymbol{u_p^*}) + \boldsymbol{\Sigma}(\boldsymbol{u_p^*}, \boldsymbol{u_p}) \boldsymbol{\Sigma}(\boldsymbol{u_p}, \boldsymbol{u_p})^{-1} \big( \boldsymbol{Y} - \boldsymbol{\mu}(\boldsymbol{u_p}) \big), \\[4pt]
    &\;\boldsymbol{\Sigma}(\boldsymbol{u_p^*}, \boldsymbol{u_p^*}) - \boldsymbol{\Sigma}(\boldsymbol{u_p^*}, \boldsymbol{u_p}) \boldsymbol{\Sigma}(\boldsymbol{u_p}, \boldsymbol{u_p})^{-1} \boldsymbol{\Sigma}(\boldsymbol{u_p}, \boldsymbol{u_p^*})
    \bigg)
    \end{aligned}
    \label{predictive_distribution}
\end{equation}
Note that in both Eq.~\eqref{eqn:joint_predictive} and \eqref{predictive_distribution} explicit dependence on $u_p$ is shown, but we emphasize that much of the covariance structure depends on $u_p$ only through $\{u_{s_i}, \nu_{z_i}\}$ as explained following Eq.~\eqref{cov_gen}. Moreover, the specific block covariances in $\boldsymbol{\Sigma}$ are computed as detailed above and in \ref{appendix_A}

\section{Application to Shock Modeling in SiC}
In this section, the proposed physics-constrained GP model is demonstrated for predicting shock velocities and state variables for single crystal SiC-3C shocked along the \([001]\) crystallographic orientation.  
The model is trained using \(N=21\) MD shock simulations, as shown in Fig.~\ref{Us_vs_Up} and detailed in \ref{sec:MD_simulations}.

To account for the multi-wave structure of the shock propagation, three distinct GP models are constructed for the leading wave, plastic wave, and phase transformation wave. This is achieved by partitioning the data into three mutually inclusive training data sets comprised of $u_p$ and the corresponding augmented state vectors, $\boldsymbol{Y}$, for each wave. We note that the data sets share common data points because the waves merge as $u_p$ increases.
Any such data points are included in both the leading wave model and the corresponding plastic or phase transformation wave model. For example, the data point having $u_p= 2.75$ km/s in Figure~\ref{Us_vs_Up} is both the leading wave and a plastic wave.

First, the leading-wave GP model, denoted by \(\mathcal{GP}_{\mathrm{lead}}\), is trained on all data points shown in blue in Figure~\ref{Us_vs_Up}. Next, the plastic wave GP model, denoted by \(\mathcal{GP}_{\mathrm{pl}}\), is trained on the set of red data points (corresponding to the trailing plastic wave) plus the subset of blue data points having $u_p>2.25$ km/s for which the leading wave is no longer elastic. Finally, the phase transformation GP model, denoted by \(\mathcal{GP}_{\mathrm{pt}}\), is trained on the set of green data points (corresponding to the trailing phase transformation wave) plus the subset of blue data points having $u_p>4.25$ km/s corresponding to the overdriven region where a phase transformation occurs in the leading wave.

The three GP models are trained sequentially to account for the multi-wave structure using the generalized (multi-wave) Hugoniot equations in Eq.~\eqref{eqn:gen_hugoniot_density}, Eq.~\eqref{Hugoniot_pressure_gen} and \eqref{Hugoniot_energy_gen}. First, \(\mathcal{GP}_{\mathrm{lead}}\) is trained such that \({\nu}_{z_{i-1}}\), \({\rho}_{i-1}\), \({P}_{i-1}\), \({T}_{i-1}\), \({E}_{i-1}\) take values associated with the ambient material state ahead of the shock \({\nu}_{z_{0}}=0\), \({\rho}_{0}\), \({P}_{0}\), \({T}_{0}\), \({E}_{0}\). Predictions are made for $u_{s_1}$ and the state variables \({\nu}_{z_{1}}\), \({\rho}_{1}\), \({P}_{1}\), \({T}_{1}\), \({E}_{1}\) in the material behind the leading wave. 

Next, \(\mathcal{GP}_{\mathrm{pl}}\) is trained such that, when the plastic wave trails a leading elastic wave, \(\mathcal{GP}_{\mathrm{lead}}\) is used to predict the material state ahead of the wave. That is, \({\nu}_{z_{i-1}}\), \({\rho}_{i-1}\), \({P}_{i-1}\), \({T}_{i-1}\), \({E}_{i-1}\) take the values of \({\nu}_{z_{1}}\), \({\rho}_{1}\), \({P}_{1}\), \({T}_{1}\), \({E}_{1}\) predicted from \(\mathcal{GP}_{\mathrm{lead}}\). When the plastic wave is the leading wave, the state variables ahead of the wave once again take values associated with the ambient material state \({\nu}_{z_{0}}=0\), \({\rho}_{0}\), \({P}_{0}\), \({T}_{0}\), \({E}_{0}\). Predictions are then made for $u_{s_i}$ and the state variables \({\nu}_{z_{i}}\), \({\rho}_{i}\), \({P}_{i}\), \({T}_{i}\), \({E}_{i}\) in the material behind the plastic wave, where $i=2$ in a multi-wave structure when the plastic wave trails and $i=1$ when the plastic wave leads.

Finally, the same procedure is followed for training \(\mathcal{GP}_{\mathrm{pt}}\) such that predictions from \(\mathcal{GP}_{\mathrm{pl}}\) are used for the material state ahead of the wave when it trails the plastic wave and the ambient state is applied ahead of the wave in the overdriven regime. 

\begin{figure}[!ht]
  \centering
  \includegraphics[width=0.8\linewidth]{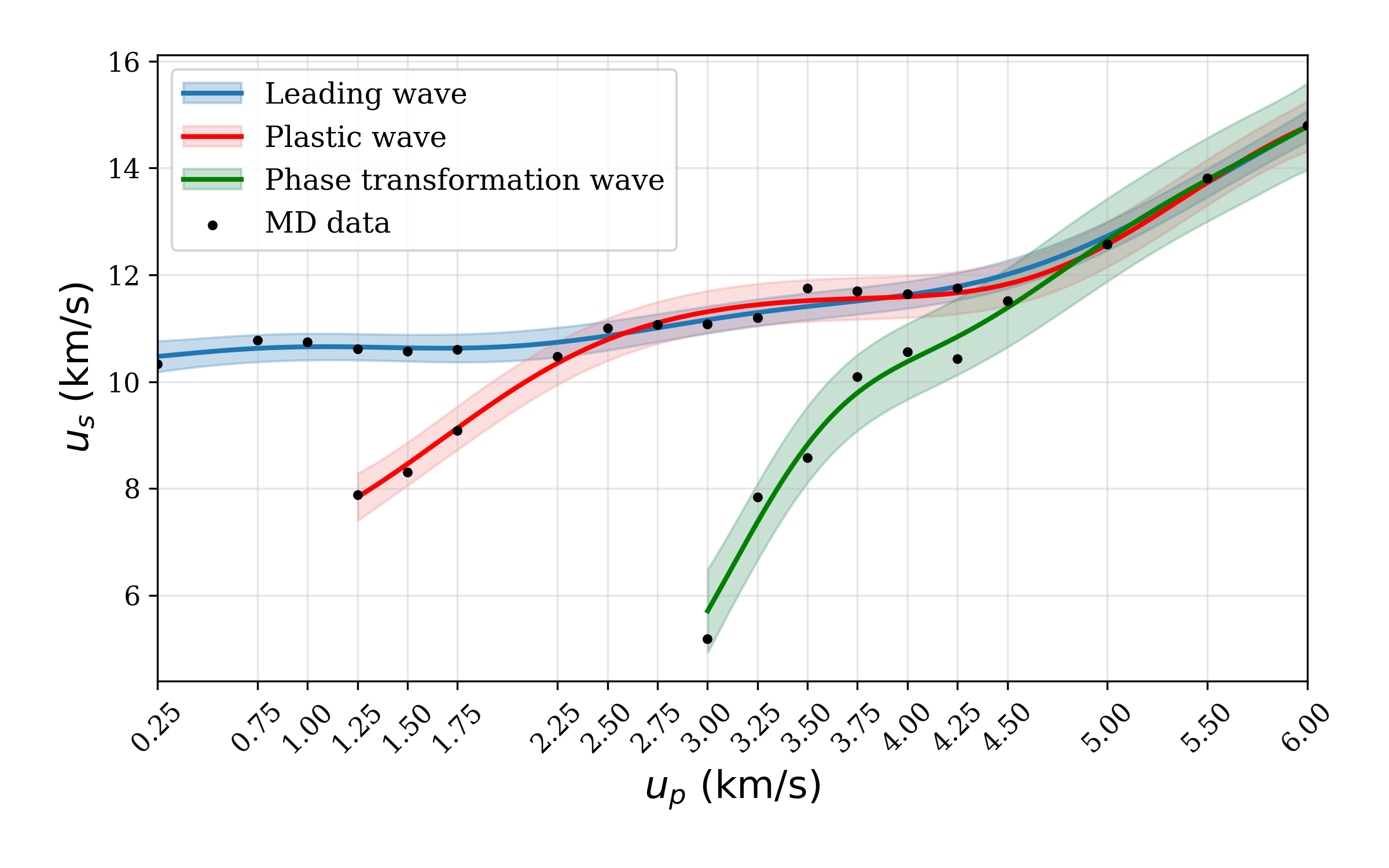}
  \caption{GP predictions of the $u_s-u_p$ relationships for the leading (blue), plastic (red), and phase transition (green) waves. The solid lines represent the mean values and the shaded regions indicate the $95\%$confidence intervals. }
  \label{fig:allwaves_Us}
\end{figure}

\begin{figure}[!ht]
  \centering
  \includegraphics[width=0.8\linewidth]{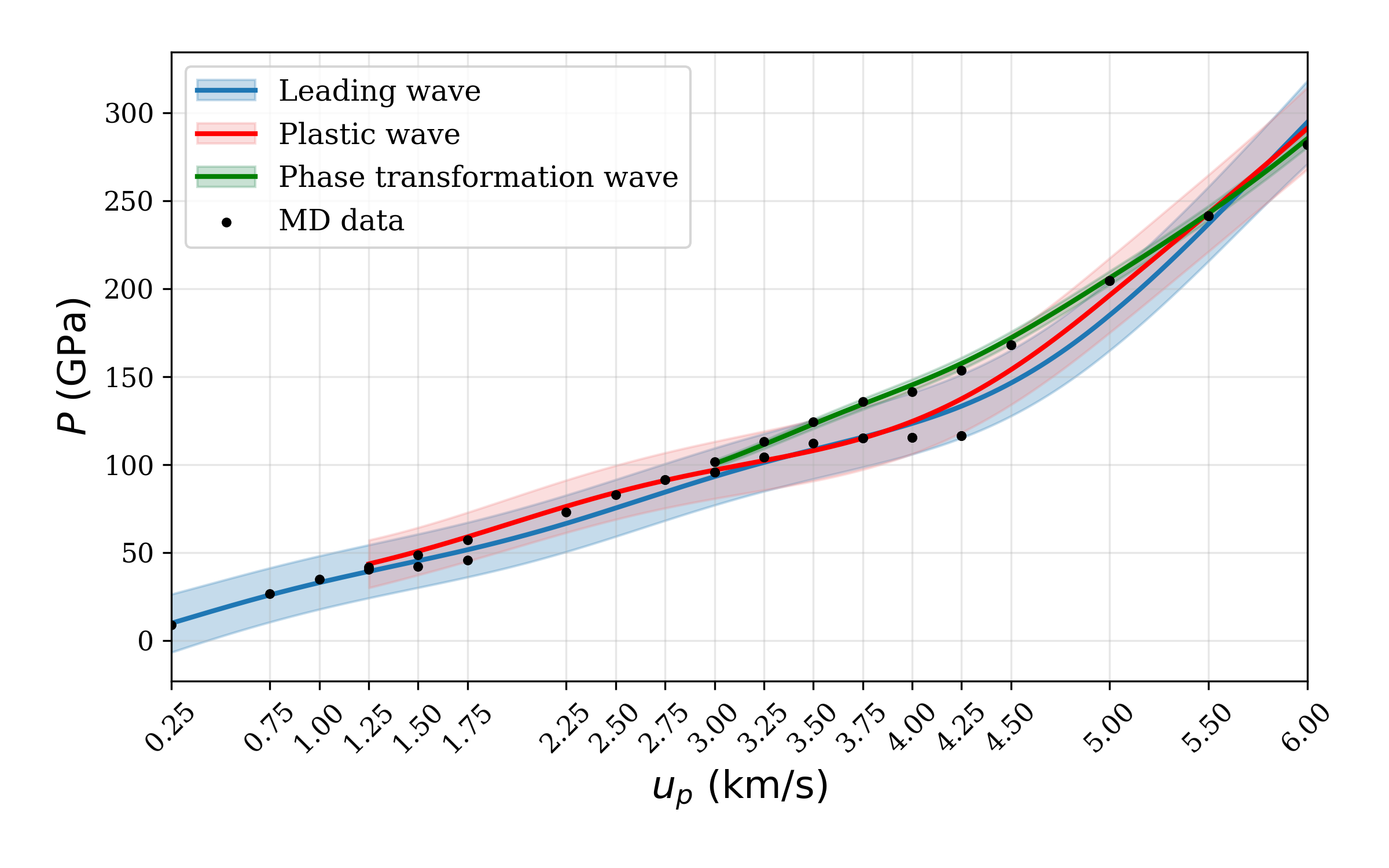}
  \caption{GP predictions for pressure $P$ in the material trailing each wave for the leading (blue), plastic (red), and phase transition (green) waves. The solid lines represent the mean values and the shaded regions indicate the $95\%$ confidence intervals.}
  \label{fig:allwaves_Pzz}
\end{figure}

\begin{figure}[!ht]
  \centering
  \includegraphics[width=0.8\linewidth]{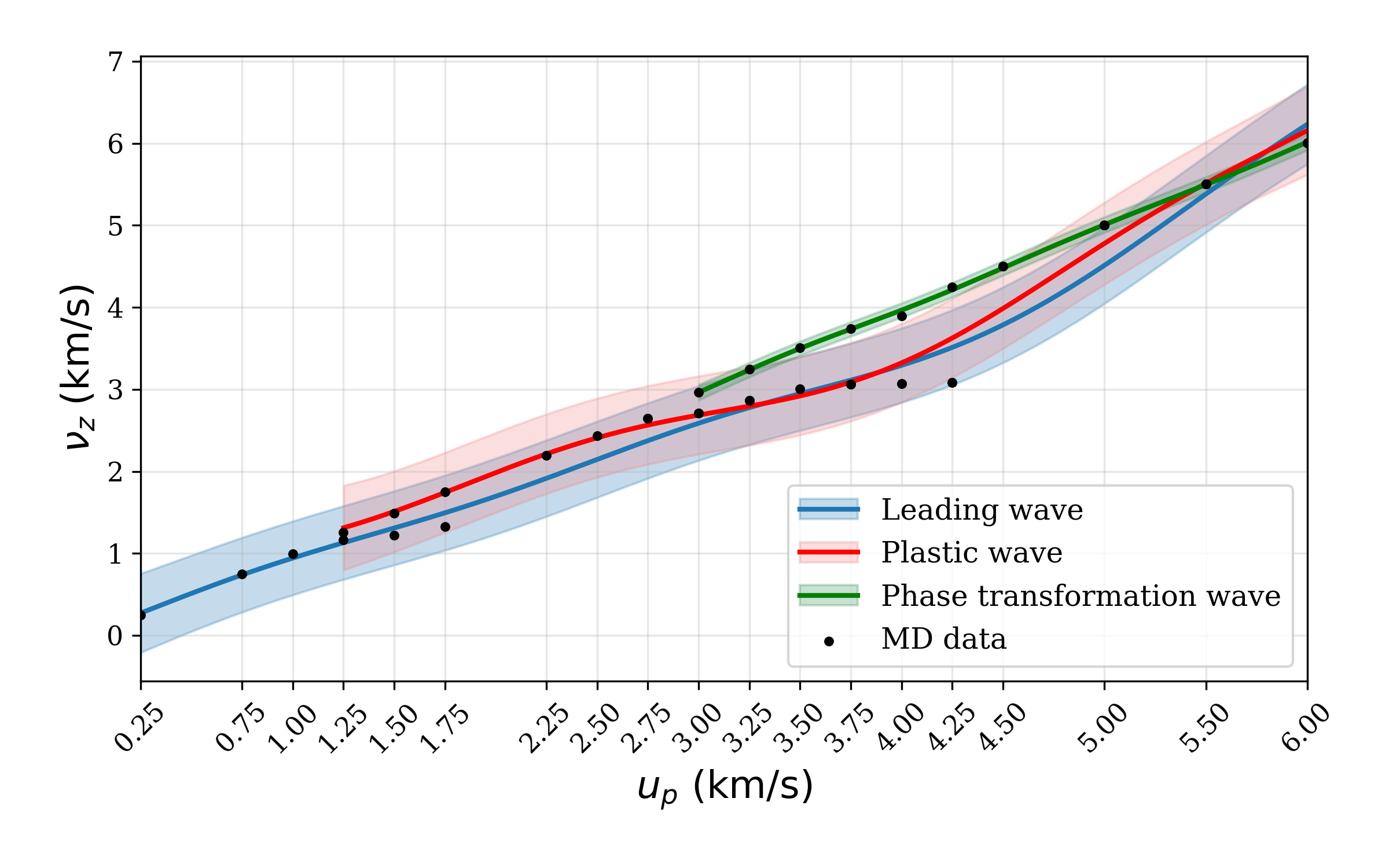}
  \caption{GP predictions for particle velocity $\nu_z$, in the material trailing each wave for the leading (blue), plastic (red), and phase transition (green) waves. The solid lines represent the mean values and the shaded regions indicate the $95$\% confidence intervals.}
  \label{fig:allwaves_Vz}
\end{figure}

\begin{figure}[!ht]
  \centering
  \includegraphics[width=0.8\linewidth]{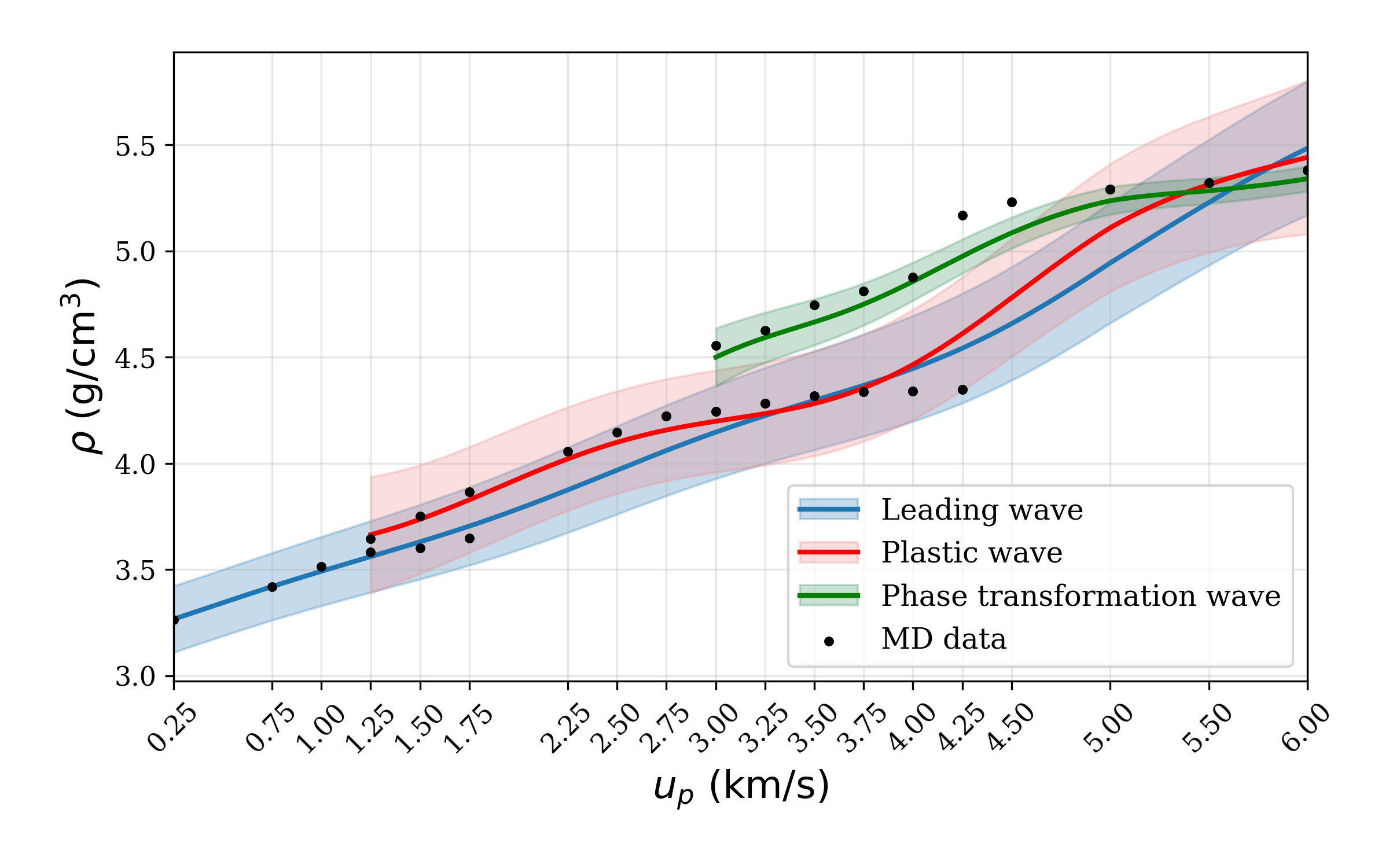}
  \caption{GP predictions for density, $\rho$ in the material trailing each wave for the leading (blue), plastic (red), and phase transition (green) waves. The solid lines represent the mean values and the shaded regions indicate the $95$\% confidence intervals.}
  \label{fig:allwaves_rho}
\end{figure}

\begin{figure}[!ht]
  \centering
  \includegraphics[width=0.8\linewidth]{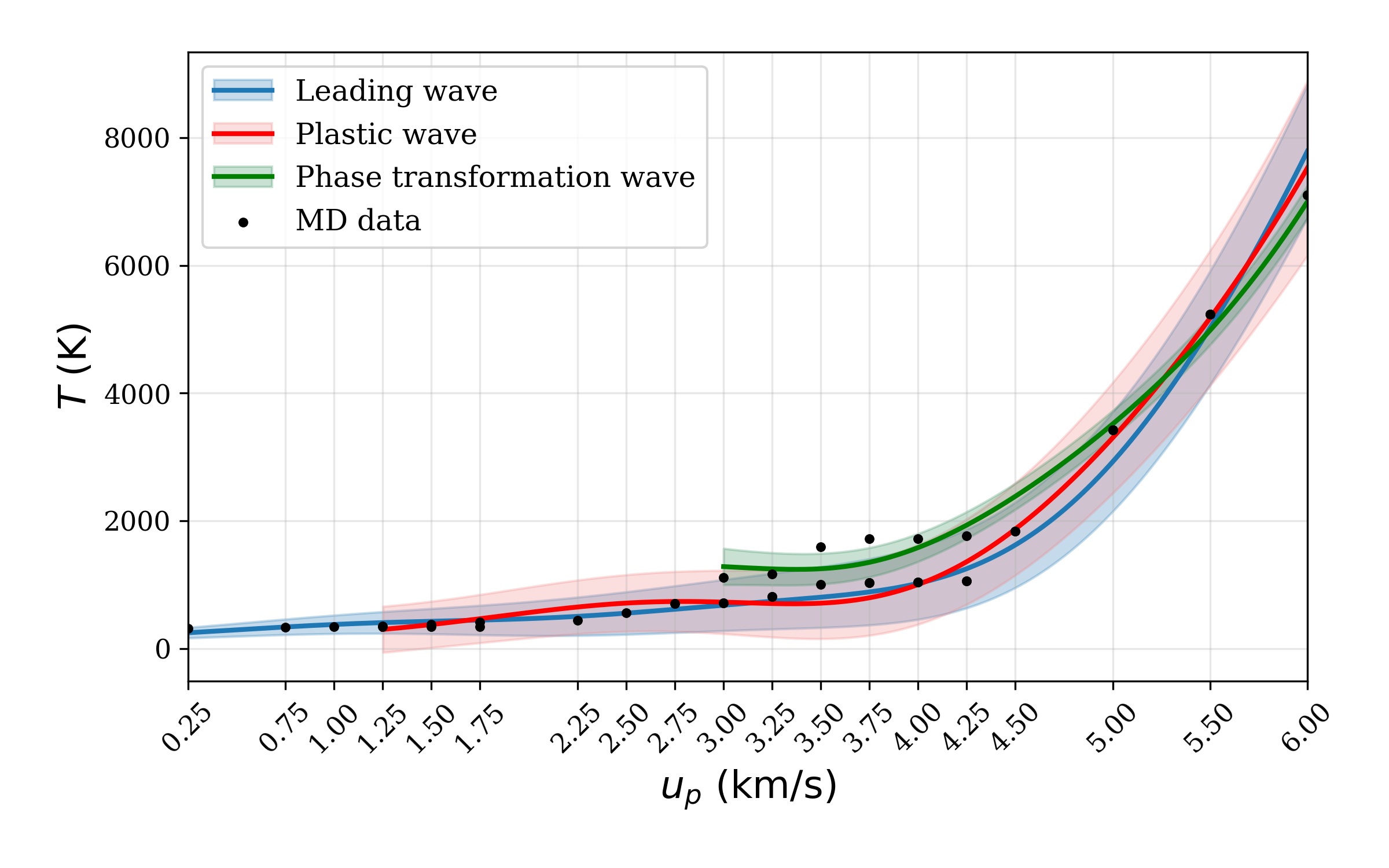}
  \caption{GP predictions for temperature, $T$ in the material trailing each wave for the leading (blue), plastic (red), and phase transition (green) waves. The solid lines represent the mean values and the shaded regions indicate the 95\% confidence intervals.}
  \label{fig:allwaves_T}
\end{figure}

The resulting GP predictions are shown in Figures~\ref{fig:allwaves_Us}--\ref{fig:allwaves_T}. Figure~\ref{fig:allwaves_Us} shows the GP predictions for each shock wave velocity, $u_s$, where the blue curve shows the leading wave with $95\%$ confidence interval, the red curve shows the plastic wave with $95\%$ confidence interval, and the green curve shows the phase transformation wave with $95\%$ confidence interval. Notice that, as expected, the leading (elastic) and plastic shock velocities converge near $u_p=2.5$ km/s. Similarly, the leading (plastic) and phase transformations converge near $u_p=4.5$ km/s. However, the uncertainties associated with each shock velocity differ. For example, the phase transformation wave has considerably higher prediction uncertainty, which implies that it is more difficult to measure from MD simulations (having higher noise) and that there is a higher degree of uncertainty in predicting when the material will reach an overdriven state (i.e. when the phase transformation wave becomes the leading wave). 

Next, Figure~\ref{fig:allwaves_Pzz} shows the GP prediction for the pressure trailing each wave. Notice here that the pressure trailing the leading wave drops slightly when a trailing (plastic or phase transformation) wave exists, but rises again when the waves merge. The GP model accurately captures this pressure drop as we see the mean leading wave GP has lower pressure than the mean plastic and phase transitions wave GPs when a dual wave structure exists ($u_p\in[1.25,2.25]$km/s and $u_p\in [3, 4.25]$ km/s), albeit with significant uncertainties.

Figure~\ref{fig:allwaves_Vz} shows the GP predicted particle velocities, $\nu_z$ for each wave. Notice that the trailing wave particle velocity tracks along $\nu_z=u_p$, because the velocity trailing the final wave must be equal to the piston velocity. However, the particle velocity ahead of the trailing wave must be less than the piston velocity. Therefore, we see that $v_z$ for the leading wave is less than $u_p$, following a similar trend to the pressure and again captured accurately by the GP model, but introducing significant uncertainty. 

Figures~\ref{fig:allwaves_rho} and \ref{fig:allwaves_T} show the GP predicted density and temperatures following each wave. Again, we see that density is lower behind the leading wave, and indeed plateaus at a particular value for the training plastic and phase transformation waves. This further implies a sharp increase in trailing density when the trailing and leading waves merge, which the GP struggles to capture. This is expected as the GP does not adapt well to sudden non-stationary features, instead driving the model toward a smooth solution with large uncertainty. 
The temperature GP meanwhile shows expected trends. It increases gradually for lower particle velocities with relatively little heat generated from elastic and plastic deformation. However, the phase transformation wave creates significant heat, such that the temperature behind the trailing phase transformation wave is consistently higher than the leading plastic wave. Once in the overdrive regime, heat generation becomes a dominant mechanism for energy dissipation and temperature increases rapidly. Even under the simple temperature assumptions from Eq.~\eqref{energy_temp_poly}, the temperature GP accurately represents these phenomena.

\begin{figure}[!ht]
  \centering
  \includegraphics[width=0.8\linewidth]{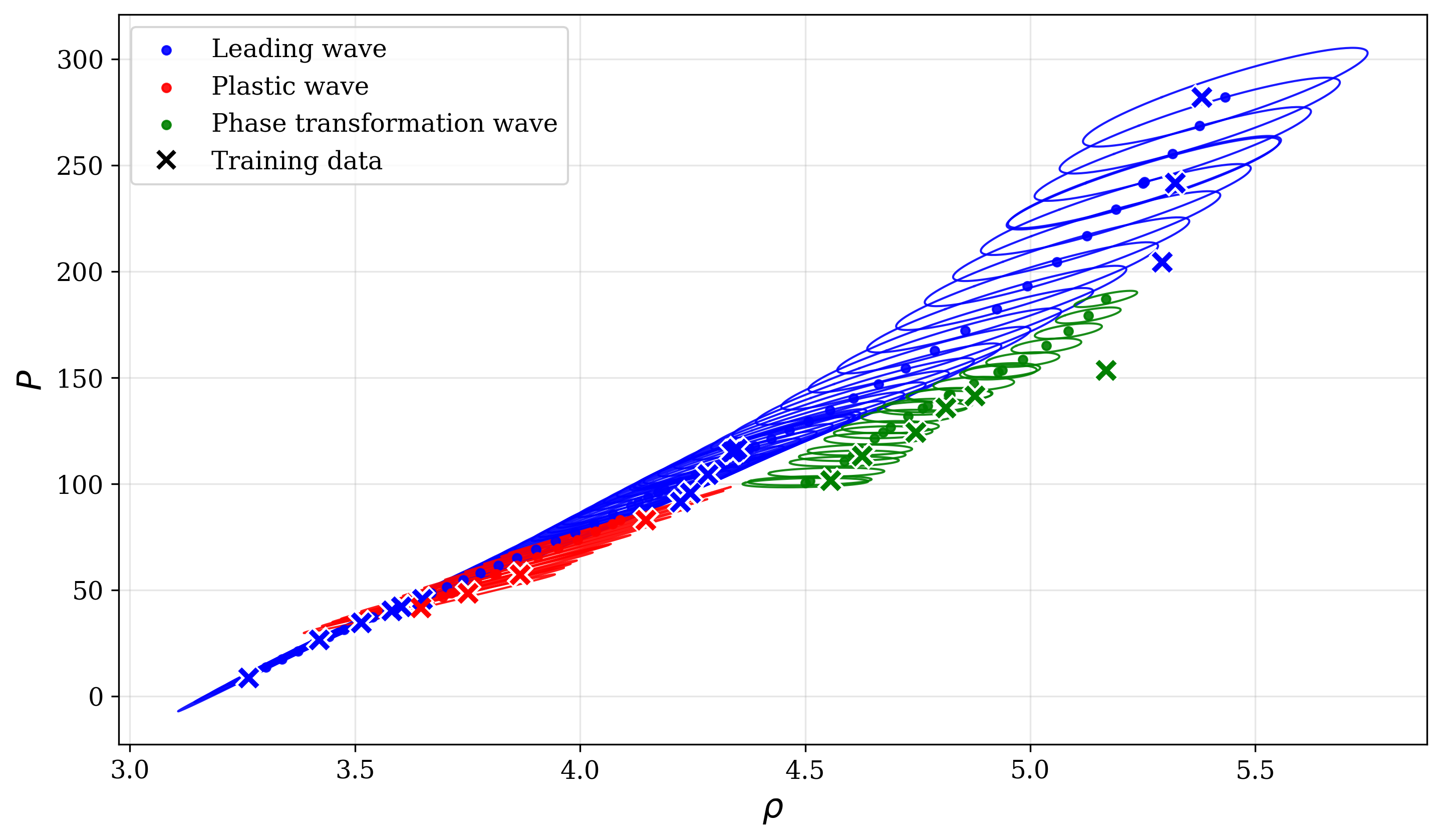}
  \caption{Predicted pressure-density Hugoniot states based on the proposed GP model. The circular markers denote the mean predictions and the ellipses represent $\pm 2 \ \mathrm{std}$ constant-probability covariance contours. The crosses indicate the training data used for each wave regime.}
  \label{P_rho}
\end{figure}

\begin{figure}[!ht]
  \centering
  \includegraphics[width=0.8\linewidth]{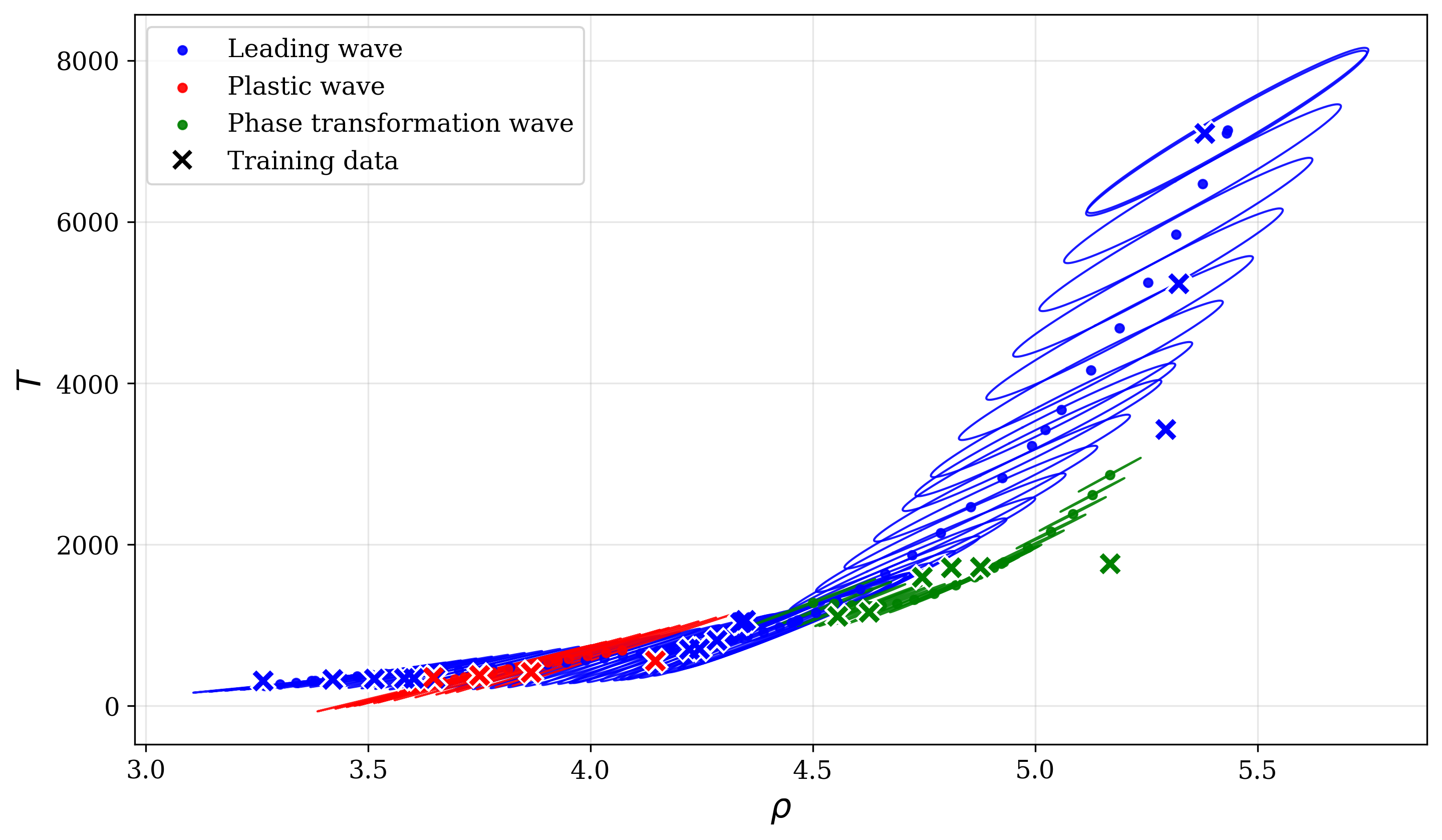}
  \caption{Predicted temperature-density Hugoniot states based on the proposed GP model. The circular markers denote the mean predictions and the ellipses represent $\pm 2 \ \mathrm{std}$ constant-probability covariance contours. The crosses indicate the training data used for each wave regime.}
  \label{T_rho}
\end{figure}

Finally, Figures~\ref{P_rho} and~\ref{T_rho} illustrate the Gaussian Process predictions of the pressure–density and temperature–density Hugoniot states across the elastic, plastic, and phase transformation regimes. We note that the Hugoniot is typically presented in terms of either volume ($V$) or inverse density ($1/\rho$). The current formulation makes presentation in this way challenging because $1/\rho$, where $\rho$ is Gaussian, follows a strongly non-Gaussian distribution with very heavy tails and undefined mean and variance. To avoid this issue, the GP could be formulated in terms of volume rather than temperature, but providing an alternate formulation is beyond our scope. Nonetheless, we observe the following expected trends. The Hugoniot is a locus of points that results from shocks a different strengths. As such, the data in Figures~\ref{P_rho} and~\ref{T_rho} are presented such that the mean predictions are shown as dots and the associated uncertainties are shown with $\pm 2,\mathrm{std}$ covariance ellipses at each discrete point. In this way, we see that the locus of points that represent the leading wave form a surface whose uncertainty increases for larger shock strengths that generate at higher $\{P, \rho\}$ states. Moreover, the trailing wave states fall below those of the leading wave states, as expected. Interestingly, the covariance ellipses show strong correlation between $\rho$ and $P$ as well as $\rho$ and $T$. Again, this is expected as even with the large uncertainties in the predictions, we expect that the values of $\rho$ and $P, T$ will increase together. 

\section{Concluding remarks}
Progress in shock physics has long relied on experimentation and modeling. Plate-impact and laser-driven experiments are standard tools for characterizing shock behavior and enabling insights into extreme phenomena that are far removed from quasi-static loading. Such experiments are, however, constrained by technical limitations related to the observation of extremely fast scales. Molecular dynamics simulations have extended this frontier by uncovering atomistic mechanisms of deformation and transformation, but their substantial computational demands restrict their broader use. To overcome these challenges and provide critical insights into the thermodynamic state of shocked materials from sparse data, the present work introduces a physics-constrained Gaussian Process (GP) regression framework that embeds the Rankine–Hugoniot jump conditions directly into the covariance structure. Relying on relatively small number of MD data, the framework yields thermodynamically consistent Hugoniot states and captures the transitions across elastic, plastic, and overdriven regimes with quantifiable uncertainty.

The proposed approach aligns with the broader goal of autonomous and data-driven materials discovery under extreme conditions. By embedding thermodynamics laws into the proposed probabilistic surrogate in conjunction with active learning strategies, it becomes possible to establish a closed-loop process in which simulations and experiments are chosen strategically to maximize information gain. Such an approach echoes recent advances in autonomous computational and physical experimentation, where Gaussian Processes have been used to navigate high-dimensional parameter spaces while accounting for inhomogeneous noise. Extending these concepts to MD simulations in shock physics can potentially drastically reduce costly trial-and-error efforts in materials understanding and discovery.

\section{Declaration of Competing Interest}

The authors declare no competing interests.

\section{Acknowledgements}

Research was sponsored by the Army Research Laboratory and was accomplished under Cooperative Agreement Number W911NF-22-2-0121. The views and conclusions contained in this document are those of the authors and should not be interpreted as representing the official policies, either expressed or implied, of the Army Research Laboratory or the U.S. Government. The U.S. Government is authorized to reproduce and distribute reprints for Government purposes notwithstanding any copyright notation herein. The authors additionally recognize the support of the Defense Threat Reduction Agency through the Materials Science in Extreme Environments Program, Award No.~HDTRA12020001.

\FloatBarrier
\appendix
\section{Covariance derivation for the block covariance model}
\label{appendix_A}
\noindent The block covariance model in Eq.~\eqref{cov_gen} can be derived as follows. Beginning with the covariance given by
\begin{equation}
    \begin{aligned}
    \mathrm{Cov}\bigl(y_l^j, y_m^k\bigr)
    &=\;
    \mathbb{E}\Bigl[\bigl(y_l^j - \mathbb{E}[y_l^j]\bigr)
                   \bigl(y_m^k - \mathbb{E}[y_m^k]\bigr)\Bigr]
    \\
    &=\;
    \mathbb{E}\bigl[y_l^j y_m^k\bigr]
    \;-\;
    \mathbb{E}[y_l^j] \mathbb{E}[y_m^k]
    \end{aligned}
    \label{Covariance_known_yjyk}
\end{equation}
we define:
\begin{equation}
    \begin{aligned}
        \mathbf{I_1}&=\mathbb{E}\bigl[y_l^jy_m^{k}\bigr]\\
        \mathbf{I_2}&=\mathbb{E}\bigl[y_l^j\bigr]\;\mathbb{E}\bigl[y_m^{k}\bigr].
    \end{aligned}
\end{equation}

\noindent 
Next, we group the terms as follows:\\

\noindent
\textit{Zeroth-order term}
\begin{equation}
    A_l^j = f_l\bigl(\bar{x}_j\bigr)
    \label{eq:zeroth_order_reduced}
\end{equation}

\noindent
\textit{First-order term}
\begin{equation}
    \begin{aligned}
        B_l^j &= 
        \left.\frac{\partial f_l}{\partial u_{s_i}}\right|_{\bar{x}_j}
        \bigl(u_{s_i}^j - \bar{u}_{s_i}^j\bigr)
        +
        \left.\frac{\partial f_l}{\partial \nu_{z_i}}\right|_{\bar{x}_j}
        \bigl(\nu_{z_i}^j - \bar{\nu}_{z_i}^j\bigr)
    \end{aligned}
    \label{eq:first_order_reduced}
\end{equation}

\noindent
\textit{Second-order term}
\begin{equation}
    \begin{aligned}
        C_l^j &= 
        \frac{1}{2}
        \left.\frac{\partial^2 f_l}{\partial (u_{s_i})^2}\right|_{\bar{x}_j}
        \bigl(u_{s_i}^j - \bar{u}_{s_i}^j\bigr)^2
        +
        \frac{1}{2}
        \left.\frac{\partial^2 f_l}{\partial (\nu_{z_i})^2}\right|_{\bar{x}_j}
        \bigl(\nu_{z_i}^j - \bar{\nu}_{z_i}^j\bigr)^2 \\
        &\quad+
        \left.\frac{\partial^2 f_l}{\partial u_{s_i} \partial \nu_{z_i}}\right|_{\bar{x}_j}
        \bigl(u_{s_i}^j - \bar{u}_{s_i}^j\bigr)
        \bigl(\nu_{z_i}^j - \bar{\nu}_{z_i}^j\bigr)
    \end{aligned}
    \label{eq:second_order_reduced}
\end{equation}

\noindent
Similarly,
\begin{equation}
    \begin{aligned}
        y_m^{k}
        &=
        A_m^{k}
        \;+\;
        B_m^{k}
        \;+\;
        C_m^{k}
        \;+\;\dots
    \end{aligned}
    \label{expansion_m_reduced}
\end{equation}

\noindent
Multiplying out and neglecting higher-order interactions yields:
\begin{equation}
    \begin{aligned}
        y_l^jy_m^{k}
        &=A_l^jA_m^{k} + A_l^jB_m^{k} + A_l^jC_m^{k} \\
        &\quad + B_l^jA_m^{k} + B_l^jB_m^{k} + B_l^jC_m^{k}\\
        &\quad + C_l^jA_m^{k} + C_l^jB_m^{k} + C_l^jC_m^{k} + \cdots
    \end{aligned}
    \label{product_expansion_reduced}
\end{equation}

\noindent
Taking expectations, using \( \mathbb{E}[B_l^j] = 0 \), we get:
\begin{equation}
    \begin{aligned}
        \mathbf{I_1} &= 
        A_l^j A_m^{k}
        \;+\;
        A_l^j\mathbb{E}[C_m^{k}]
        \;+\;
        \mathbb{E}[B_l^j B_m^{k}]
        \;+\;
        \mathbb{E}[C_l^j] A_m^{k}
    \end{aligned}
    \label{eq:I1_reduced}
\end{equation}

\begin{equation}
    \begin{aligned}
        \mathbf{I_2} &= \mathbb{E}[y_l^j]  \mathbb{E}[y_m^{k}] \\
        &= A_l^j A_m^{k} + A_l^j \mathbb{E}[C_m^{k}] + \mathbb{E}[C_l^j] A_m^{k} + \mathbb{E}[C_l^j] \mathbb{E}[C_m^{k}]
    \end{aligned}
    \label{eq:I2_reduced}
\end{equation}

\noindent
Subtracting:
\begin{equation}
    \mathrm{Cov}\bigl(y_l^j,y_m^{k}\bigr)
    = \mathbb{E}[B_l^j B_m^{k}] - \mathbb{E}[C_l^j] \mathbb{E}[C_m^{k}]
\end{equation}

\noindent
If we neglect the \(\mathbb{E}[C_l^j] \mathbb{E}[C_m^{k}]\) term as higher-order, we obtain:
\begin{equation}
    \begin{aligned}
        \mathrm{Cov}\bigl(y_l^j,y_m^{k}\bigr) \approx \mathbb{E}[B_l^j B_m^{k}] =\;&
    \left.\frac{\partial f_l}{\partial u_{s_i}}\right|_{\bar{x}_j}
    \left.\frac{\partial f_m}{\partial u_{s_i}}\right|_{\bar{x}_k} \mathbf{K}_{u_{s_i}u_{s_i}}^{jk} \\
    &+
    \left(
    \left.\frac{\partial f_l}{\partial u_{s_i}}\right|_{\bar{x}_j}
    \left.\frac{\partial f_m}{\partial \nu_{z_i}}\right|_{\bar{x}_k}
    +
    \left.\frac{\partial f_l}{\partial \nu_{z_i}}\right|_{\bar{x}_j}
    \left.\frac{\partial f_m}{\partial u_{s_i}}\right|_{\bar{x}_k}
    \right) \cdot \mathbf{K}_{u_{s_i}\nu_{z_i}}^{jk} \\
    &+
    \left.\frac{\partial f_l}{\partial \nu_{z_i}}\right|_{\bar{x}_j}
    \left.\frac{\partial f_m}{\partial \nu_{z_i}}\right|_{\bar{x}_k}
    \mathbf{K}_{\nu_{z_i}\nu_{z_i}}^{jk}
    \end{aligned}
    \label{eq:covariance_cleaned}
\end{equation}

\section{Derivation of the Hugoniot-based derivatives}\label{appendix_B}
\noindent Considering the Hugoniot equations where indices \(i\) and \(j\) represent the material and the state, respectively. The Hugoniot equations are given as:
\begin{equation}\label{Hugoniot_red1}
    P_i^{j}=
    \rho_{i-1}^{j}\bigl(u_{s_i}^{j}-\nu_{z_{i-1}}^{j}\bigr)
    \bigl(\nu_{z_i}^{j}-\nu_{z_{i-1}}^{j}\bigr)
    + P_{i-1}^{j}
\end{equation}

\begin{equation}\label{Hugoniot_red2}
    \rho_i^{j} =
    \rho_{i-1}^{j}
    \frac{u_{s_i}^{j}-\nu_{z_{i-1}}^{j}}
         {u_{s_i}^{j}-\nu_{z_i}^{j}}.
\end{equation}

\begin{equation}\label{Hugoniot_red3}
    E_i^{j} = E_{i-1}^{j}
    +\frac12\bigl(P_i^{j}+P_{i-1}^{j}\bigr)
    \left(\frac{1}{\rho_{i-1}^{j}}-\frac{1}{\rho_i^{j}}\right)
    \end{equation}
The varrious derivatives are derived as follows.

\noindent
\textit{Pressure Derivatives}
\begin{align}
    \frac{\partial P_i^j}{\partial u_{s_i}^j} 
    &= \rho_{i-1}^j(\nu_{z_i}^j - \nu_{z_{i-1}}^j)
    \label{dP_du} \\[3pt]
    \frac{\partial P_i^j}{\partial \nu_{z_i}^j} 
    &= \rho_{i-1}^j(u_{s_i}^j - \nu_{z_{i-1}}^j)
    \label{dP_dv} \\[3pt]
    \frac{\partial^2 P_i^j}{\partial (u_{s_i}^j)^2} &= 0 
    \label{d2P_du2} \\[3pt]
    \frac{\partial^2 P_i^j}{\partial (\nu_{z_i}^j)^2} &= 0 
    \label{d2P_dv2} \\[3pt]
    \frac{\partial^2 P_i^j}{\partial u_{s_i}^j \partial \nu_{z_i}^j} 
    &= \rho_{i-1}^j
    \label{d2P_du_dv}
\end{align}

\noindent
\textit{Density Derivatives}
\begin{align}
    \frac{\partial \rho_i^j}{\partial u_{s_i}^j} 
    &= \rho_{i-1}^j \frac{\nu_{z_i}^j - \nu_{z_{i-1}}^j}{(u_{s_i}^j - \nu_{z_i}^j)^2}
    \label{drho_du} \\[3pt]
    \frac{\partial \rho_i^j}{\partial \nu_{z_i}^j} 
    &= \rho_{i-1}^j \frac{u_{s_i}^j - \nu_{z_{i-1}}^j}{(u_{s_i}^j - \nu_{z_i}^j)^2}
    \label{drho_dv} \\[6pt]
    \frac{\partial^2 \rho_i^j}{\partial (u_{s_i}^j)^2} 
    &= 2\rho_{i-1}^j \frac{\nu_{z_{i-1}}^j - \nu_{z_i}^j}{(u_{s_i}^j - \nu_{z_i}^j)^3}
    \label{d2rho_du2} \\[3pt]
    \frac{\partial^2 \rho_i^j}{\partial (\nu_{z_i}^j)^2} 
    &= 2\rho_{i-1}^j \frac{u_{s_i}^j - \nu_{z_{i-1}}^j}{(u_{s_i}^j - \nu_{z_i}^j)^3}
    \label{d2rho_dv2} \\[3pt]
    \frac{\partial^2 \rho_i^j}{\partial u_{s_i}^j \partial \nu_{z_i}^j} 
    &= -2\rho_{i-1}^j \frac{u_{s_i}^j - \nu_{z_{i-1}}^j}{(u_{s_i}^j - \nu_{z_i}^j)^3}
    \label{d2rho_du_dv}
\end{align}

\noindent
\textit{Energy Definition}
\begin{equation}
    E_i^j = E_{i-1}^j 
          + \frac{1}{2}(\nu_{z_i}^j - \nu_{z_{i-1}}^j)^2 
          + P_{i-1}^j \cdot \frac{1}{\rho_{i-1}^j} 
            \frac{\nu_{z_i}^j - \nu_{z_{i-1}}^j}{u_{s_i}^j - \nu_{z_{i-1}}^j}
    \label{E_def}
\end{equation}

\noindent
\textit{Energy Derivatives}
\begin{align}
    \frac{\partial E_i^j}{\partial u_{s_i}^j} &=
    -\frac{P_{i-1}^j}{\rho_{i-1}^j} 
     \frac{\nu_{z_i}^j - \nu_{z_{i-1}}^j}{(u_{s_i}^j - \nu_{z_{i-1}}^j)^2}
    \label{dE_du} \\[3pt]
    \frac{\partial E_i^j}{\partial \nu_{z_i}^j} &=
    (\nu_{z_i}^j - \nu_{z_{i-1}}^j) 
    + \frac{P_{i-1}^j}{\rho_{i-1}^j(u_{s_i}^j - \nu_{z_{i-1}}^j)}
    \label{dE_dv}
\end{align}

\begin{align}
    \frac{\partial^2 E_i^j}{\partial (u_{s_i}^j)^2} &=
    2\frac{P_{i-1}^j}{\rho_{i-1}^j} 
    \frac{\nu_{z_i}^j - \nu_{z_{i-1}}^j}{(u_{s_i}^j - \nu_{z_{i-1}}^j)^3}
    \label{d2E_du2} \\[3pt]
    \frac{\partial^2 E_i^j}{\partial (\nu_{z_i}^j)^2} &= 1
    \label{d2E_dv2} \\[3pt]
    \frac{\partial^2 E_i^j}{\partial u_{s_i}^j \partial \nu_{z_i}^j} &=
    -\frac{P_{i-1}^j}{\rho_{i-1}^j} 
    \frac{1}{(u_{s_i}^j - \nu_{z_{i-1}}^j)^2}
    \label{d2E_du_dv}
\end{align}

\noindent
\textit{Temperature derivatives}
\begin{equation}\label{eq:T-model}
  T_i^{j}=a+bE_i^{j}
\end{equation}

\begin{equation}\label{eq:dT-dus}
  \frac{\partial T_i^{j}}{\partial u_{s_i}^{j}}
  =
  b\frac{\partial E_i^{j}}{\partial u_{s_i}^{j}} 
\end{equation}

\begin{equation}\label{eq:dT-dvz}
  \frac{\partial T_i^{j}}{\partial \nu_{z_i}^{j}}
  =
  b\frac{\partial E_i^{j}}{\partial \nu_{z_i}^{j}}
\end{equation}

\begin{equation}\label{eq:d2T-dus2}
  \frac{\partial^2 T_i^{j}}{\partial (u_{s_i}^{j})^2}
  =
  b\frac{\partial^2 E_i^{j}}{\partial (u_{s_i}^{j})^2}
\end{equation}

\begin{equation}\label{eq:d2T-dvz2}
  \frac{\partial^2 T_i^{j}}{\partial (\nu_{z_i}^{j})^2}
  =
  b\frac{\partial^2 E_i^{j}}{\partial (\nu_{z_i}^{j})^2}
\end{equation}

\begin{equation}\label{eq:d2T-dus-dvz}
  \frac{\partial^2 T_i^{j}}{\partial u_{s_i}^{j}\partial \nu_{z_i}^{j}}
  =
  b\frac{\partial^2 E_i^{j}}{\partial u_{s_i}^{j}\partial \nu_{z_i}^{j}}
\end{equation}

\section{Mean and covariances for Thermodynamic State Variables}\label{appendix_C}
Consider the Taylor series expansion mean and covariance expressions in Eqs.~\eqref{Taylor_mean_gen} and \eqref{Covariance_yjyk_final}, respectively. Using the analytical derivatives in \ref{appendix_B}, the explicit expression of the thermodynamic variables are given below. The index \(i\) denotes the propagating wave front, and \(j\) and \(k\) are the data points of the \(l^{th}\) and \(m^{th}\) output variables.

\begin{equation}
P_i^j = \rho_{i-1}^j\big(u_{s_i}^j-\nu_{z_{i-1}}^j\big)\big(\nu_{z_i}^j-\nu_{z_{i-1}}^j\big)+P_{i-1}^j
\end{equation}

\noindent
\textit{Mean pressure}\\
Using Eq.~\eqref{Hugoniot_red1} 
\begin{equation}
\bar P_i^{j} =
\rho_{i-1}^{j}
\big(\bar u_{s_i}^{j}-\nu_{z_{i-1}}^{j}\big)
\big(\bar \nu_{z_i}^{j}-\nu_{z_{i-1}}^{j}\big)
+ P_{i-1}^{j}
\;+\;
\rho_{i-1}^{j}\boldsymbol{K}_{u_{s_i}\nu_{z_i}}^{jj}
\end{equation}

\noindent
\textit{Pressure covariances}\\
\noindent Using the derivatives in Eqs.~\eqref{dP_du}-\eqref{d2P_du_dv} yields 
\begin{equation}
\begin{aligned}
\boldsymbol{K}_{P_iP_i}^{jk}
&=
\rho_{i-1}^{j}\big(\bar \nu_{z_i}^{j}-\nu_{z_{i-1}}^{j}\big)
\rho_{i-1}^{k}\big(\bar \nu_{z_i}^{k}-\nu_{z_{i-1}}^{k}\big)\boldsymbol{K}_{u_{s_i}u_{s_i}}^{jk}
\\
&\quad+
\Big[
\rho_{i-1}^{j}\big(\bar \nu_{z_i}^{j}-\nu_{z_{i-1}}^{j}\big)
\rho_{i-1}^{k}\big(\bar u_{s_i}^{k}-\nu_{z_{i-1}}^{k}\big)
+
\rho_{i-1}^{j}\big(\bar u_{s_i}^{j}-\nu_{z_{i-1}}^{j}\big)
\rho_{i-1}^{k}\big(\bar \nu_{z_i}^{k}-\nu_{z_{i-1}}^{k}\big)
\Big]\boldsymbol{K}_{u_{s_i}\nu_{z_i}}^{jk}
\\
&\quad+
\rho_{i-1}^{j}\big(\bar u_{s_i}^{j}-\nu_{z_{i-1}}^{j}\big)
\rho_{i-1}^{k}\big(\bar u_{s_i}^{k}-\nu_{z_{i-1}}^{k}\big)\boldsymbol{K}_{\nu_{z_i}\nu_{z_i}}^{jk}
\end{aligned}
\end{equation}

\noindent
\textit{Mean density}\\
\noindent Using Eq.~\eqref{Hugoniot_red2} gives

\begin{equation}
\label{rho_mean}
\begin{aligned}
\bar\rho_i^{j}
&=
\rho_{i-1}^{j}
\frac{\bar u_{s_i}^{j}-\nu_{z_{i-1}}^{j}}
     {\bar u_{s_i}^{j}-\bar\nu_{z_i}^{j}}
\\[4pt]
&\quad+
\rho_{i-1}^{j}
\frac{
(\nu_{z_{i-1}}^{j}-\bar\nu_{z_i}^{j})\boldsymbol{K}_{u_{s_i}u_{s_i}}^{jj}
+(\bar u_{s_i}^{j}-\nu_{z_{i-1}}^{j})\boldsymbol{K}_{\nu_{z_i}\nu_{z_i}}^{jj}
-2(\bar u_{s_i}^{j}-\nu_{z_{i-1}}^{j})\boldsymbol{K}_{u_{s_i}\nu_{z_i}}^{jj}}
     {(\bar u_{s_i}^{j}-\bar\nu_{z_i}^{j})^{3}}
\end{aligned}
\end{equation}

\noindent
\textit{Density covariances}\\
\noindent Using the derivatives in Eqs.~\eqref{drho_du}-\eqref{d2rho_du_dv} yields 
\begin{equation}
\label{K_rhorho}
\begin{aligned}
\boldsymbol{K}_{\rho_i\rho_i}^{jk}
&=
\frac{\rho_{i-1}^{j}(\bar\nu_{z_i}^{j}-\nu_{z_{i-1}}^{j})}
     {(\bar u_{s_i}^{j}-\bar\nu_{z_i}^{j})^{2}}
\frac{\rho_{i-1}^{k}(\bar\nu_{z_i}^{k}-\nu_{z_{i-1}}^{k})}
     {(\bar u_{s_i}^{k}-\bar\nu_{z_i}^{k})^{2}}
\boldsymbol{K}_{u_{s_i}u_{s_i}}^{jk}
\\
&\quad+
\Bigg[
\frac{\rho_{i-1}^{j}(\bar\nu_{z_i}^{j}-\nu_{z_{i-1}}^{j})}
     {(\bar u_{s_i}^{j}-\bar\nu_{z_i}^{j})^{2}}
\frac{\rho_{i-1}^{k}(\bar u_{s_i}^{k}-\nu_{z_{i-1}}^{k})}
     {(\bar u_{s_i}^{k}-\bar\nu_{z_i}^{k})^{2}}
\\[-2pt]
&\hspace{19mm}+
\frac{\rho_{i-1}^{j}(\bar u_{s_i}^{j}-\nu_{z_{i-1}}^{j})}
     {(\bar u_{s_i}^{j}-\bar\nu_{z_i}^{j})^{2}}
\frac{\rho_{i-1}^{k}(\bar\nu_{z_i}^{k}-\nu_{z_{i-1}}^{k})}
     {(\bar u_{s_i}^{k}-\bar\nu_{z_i}^{k})^{2}}
\Bigg]
\boldsymbol{K}_{u_{s_i}\nu_{z_i}}^{jk}
\\
&\quad+
\frac{\rho_{i-1}^{j}(\bar u_{s_i}^{j}-\nu_{z_{i-1}}^{j})}
     {(\bar u_{s_i}^{j}-\bar\nu_{z_i}^{j})^{2}}
\frac{\rho_{i-1}^{k}(\bar u_{s_i}^{k}-\nu_{z_{i-1}}^{k})}
     {(\bar u_{s_i}^{k}-\bar\nu_{z_i}^{k})^{2}}
\boldsymbol{K}_{\nu_{z_i}\nu_{z_i}}^{jk}
\end{aligned}
\end{equation}

\noindent
\textit{Mean energy}\\
\noindent Using Eqs.~\eqref{Hugoniot_red1}--\eqref{Hugoniot_red3} gives
\begin{equation}
\label{E_mean}
\begin{aligned}
\bar E_i^{j}
&=
E_{i-1}^{j}
+\tfrac12(\bar\nu_{z_i}^{j}-\nu_{z_{i-1}}^{j})^{2}
+\frac{P_{i-1}^{j}}{\rho_{i-1}^{j}}
  \frac{\bar\nu_{z_i}^{j}-\nu_{z_{i-1}}^{j}}
       {\bar u_{s_i}^{j}-\nu_{z_{i-1}}^{j}}
\\[4pt]
&\quad+
\frac12\Bigg[
\Bigl(2\tfrac{P_{i-1}^{j}}{\rho_{i-1}^{j}}
       \tfrac{\bar\nu_{z_i}^{j}-\nu_{z_{i-1}}^{j}}
             {(\bar u_{s_i}^{j}-\nu_{z_{i-1}}^{j})^{3}}\Bigr)
        \boldsymbol{K}_{u_{s_i}u_{s_i}}^{jj}
\;+\;
\boldsymbol{K}_{\nu_{z_i}\nu_{z_i}}^{jj}
\\[2pt]
&\hspace{36mm}
-2\tfrac{P_{i-1}^{j}}{\rho_{i-1}^{j}}
       \tfrac{1}{(\bar u_{s_i}^{j}-\nu_{z_{i-1}}^{j})^{2}}
       \boldsymbol{K}_{u_{s_i}\nu_{z_i}}^{jj}
\Bigg]
\end{aligned}
\end{equation}

\noindent
\textit{Energy covariances}\\
\noindent Using the derivatives in Eqs.~\eqref{dE_du}-\eqref{d2E_du_dv} yields
\begin{equation}
\label{K_EE}
\begin{aligned}
\boldsymbol{K}_{E_iE_i}^{jk}
&=
\Bigl(\!-\frac{P_{i-1}^{j}}{\rho_{i-1}^{j}}
        \frac{\bar\nu_{z_i}^{j}-\nu_{z_{i-1}}^{j}}
             {(\bar u_{s_i}^{j}-\nu_{z_{i-1}}^{j})^{2}}\Bigr)
\Bigl(\!-\frac{P_{i-1}^{k}}{\rho_{i-1}^{k}}
        \frac{\bar\nu_{z_i}^{k}-\nu_{z_{i-1}}^{k}}
             {(\bar u_{s_i}^{k}-\nu_{z_{i-1}}^{k})^{2}}\Bigr)
\boldsymbol{K}_{u_{s_i}u_{s_i}}^{jk}
\\
&\quad+
\Bigg[
-\frac{P_{i-1}^{j}}{\rho_{i-1}^{j}}
  \frac{\bar\nu_{z_i}^{j}-\nu_{z_{i-1}}^{j}}
       {(\bar u_{s_i}^{j}-\nu_{z_{i-1}}^{j})^{2}}
\Bigl(\bar\nu_{z_i}^{k}-\nu_{z_{i-1}}^{k}
      +\tfrac{P_{i-1}^{k}}{\rho_{i-1}^{k}}
       \tfrac{1}{\bar u_{s_i}^{k}-\nu_{z_{i-1}}^{k}}\Bigr)
\\[-2pt]
&\hspace{20mm}+
\Bigl(\bar\nu_{z_i}^{j}-\nu_{z_{i-1}}^{j}
      +\tfrac{P_{i-1}^{j}}{\rho_{i-1}^{j}}
       \tfrac{1}{\bar u_{s_i}^{j}-\nu_{z_{i-1}}^{j}}\Bigr)
\Bigl(\!-\frac{P_{i-1}^{k}}{\rho_{i-1}^{k}}
        \frac{\bar\nu_{z_i}^{k}-\nu_{z_{i-1}}^{k}}
             {(\bar u_{s_i}^{k}-\nu_{z_{i-1}}^{k})^{2}}\Bigr)
\Bigg]
\boldsymbol{K}_{u_{s_i}\nu_{z_i}}^{jk}
\\
&\quad+
\Bigl(\bar\nu_{z_i}^{j}-\nu_{z_{i-1}}^{j}
      +\tfrac{P_{i-1}^{j}}{\rho_{i-1}^{j}}
       \tfrac{1}{\bar u_{s_i}^{j}-\nu_{z_{i-1}}^{j}}\Bigr)
\Bigl(\bar\nu_{z_i}^{k}-\nu_{z_{i-1}}^{k}
      +\tfrac{P_{i-1}^{k}}{\rho_{i-1}^{k}}
       \tfrac{1}{\bar u_{s_i}^{k}-\nu_{z_{i-1}}^{k}}\Bigr)
\boldsymbol{K}_{\nu_{z_i}\nu_{z_i}}^{jk}
\end{aligned}
\end{equation}

\noindent
\textit{Pressure–shock velocity cross-covariance}\\
\noindent Using the derivatives in Eqs.~\eqref{dP_du}-\eqref{d2P_du_dv} one obtains
\begin{equation}
\label{K_Pus}
\boldsymbol{K}_{P_iu_{s_i}}^{jk} =
\rho_{i-1}^{j}\bigl(\bar\nu_{z_i}^{j}-\nu_{z_{i-1}}^{j}\bigr)
\boldsymbol{K}_{u_{s_i}u_{s_i}}^{jk}
\;+\;
\rho_{i-1}^{j}\bigl(\bar u_{s_i}^{j}-\nu_{z_{i-1}}^{j}\bigr)
\boldsymbol{K}_{u_{s_i}\nu_{z_i}}^{jk}
\end{equation}

\noindent
\textit{Pressure–particle velocity cross-covariance}\\
\noindent Using the derivatives in Eqs.~\eqref{dP_du}–\eqref{d2P_du_dv} leads to 
\begin{equation}
\label{K_Pvz}
\boldsymbol{K}_{P_i\nu_{z_i}}^{jk} =
\rho_{i-1}^{j}\bigl(\bar\nu_{z_i}^{j}-\nu_{z_{i-1}}^{j}\bigr)
\boldsymbol{K}_{u_{s_i}\nu_{z_i}}^{jk}
\;+\;
\rho_{i-1}^{j}\bigl(\bar u_{s_i}^{j}-\nu_{z_{i-1}}^{j}\bigr)
\boldsymbol{K}_{\nu_{z_i}\nu_{z_i}}^{jk}
\end{equation}

\noindent
\textit{Pressure-density cross covariance}\\
Using Eqs.~\eqref{dP_du}--\eqref{d2P_du_dv} and
\eqref{drho_du}--\eqref{d2rho_du_dv} yields
\begin{equation}
\label{K_Prho}
\begin{aligned}
\boldsymbol{K}_{P_i\rho_i}^{jk}
&=
\rho_{i-1}^{j}\bigl(\bar\nu_{z_i}^{j}-\nu_{z_{i-1}}^{j}\bigr)
\;
\frac{\rho_{i-1}^{k}\bigl(\bar\nu_{z_i}^{k}-\nu_{z_{i-1}}^{k}\bigr)}
     {(\bar u_{s_i}^{k}-\bar\nu_{z_i}^{k})^{2}}
\boldsymbol{K}_{u_{s_i}u_{s_i}}^{jk}
\\
&\quad+
\Bigl[
\rho_{i-1}^{j}\bigl(\bar\nu_{z_i}^{j}-\nu_{z_{i-1}}^{j}\bigr)
\;
\frac{\rho_{i-1}^{k}\bigl(\bar u_{s_i}^{k}-\nu_{z_{i-1}}^{k}\bigr)}
     {(\bar u_{s_i}^{k}-\bar\nu_{z_i}^{k})^{2}}
+
\rho_{i-1}^{j}\bigl(\bar u_{s_i}^{j}-\nu_{z_{i-1}}^{j}\bigr)
\;
\frac{\rho_{i-1}^{k}\bigl(\bar\nu_{z_i}^{k}-\nu_{z_{i-1}}^{k}\bigr)}
     {(\bar u_{s_i}^{k}-\bar\nu_{z_i}^{k})^{2}}
\Bigr]
\boldsymbol{K}_{u_{s_i}\nu_{z_i}}^{jk}
\\
&\quad+
\rho_{i-1}^{j}\bigl(\bar u_{s_i}^{j}-\nu_{z_{i-1}}^{j}\bigr)
\;
\frac{\rho_{i-1}^{k}\bigl(\bar u_{s_i}^{k}-\nu_{z_{i-1}}^{k}\bigr)}
     {(\bar u_{s_i}^{k}-\bar\nu_{z_i}^{k})^{2}}
\boldsymbol{K}_{\nu_{z_i}\nu_{z_i}}^{jk}
\end{aligned}
\end{equation}

\noindent
\textit{Pressure–energy cross-covariance}\\
\noindent Using the derivatives in Eqs.~\eqref{dP_du}–\eqref{d2P_du_dv} and \eqref{dE_du}–\eqref{d2E_du_dv} one obtains
\begin{equation}
\label{K_PE}
\begin{aligned}
\boldsymbol{K}_{P_iE_i}^{jk}
&=
\rho_{i-1}^{j}\bigl(\bar\nu_{z_i}^{j}-\nu_{z_{i-1}}^{j}\bigr)
\Bigl[-\tfrac{P_{i-1}^{k}}{\rho_{i-1}^{k}}
       \tfrac{\bar\nu_{z_i}^{k}-\nu_{z_{i-1}}^{k}}
             {(\bar u_{s_i}^{k}-\nu_{z_{i-1}}^{k})^{2}}\Bigr]
\boldsymbol{K}_{u_{s_i}u_{s_i}}^{jk}
\\
&\quad+
\Bigl[
\rho_{i-1}^{j}\bigl(\bar\nu_{z_i}^{j}-\nu_{z_{i-1}}^{j}\bigr)
\Bigl(\bar\nu_{z_i}^{k}-\nu_{z_{i-1}}^{k}
      +\tfrac{P_{i-1}^{k}}{\rho_{i-1}^{k}}
       \tfrac{1}{\bar u_{s_i}^{k}-\nu_{z_{i-1}}^{k}}\Bigr)
\\[-2pt]
&\hspace{20mm}+
\rho_{i-1}^{j}(\bar u_{s_i}^{j}-\nu_{z_{i-1}}^{j})
\Bigl[-\tfrac{P_{i-1}^{k}}{\rho_{i-1}^{k}}
       \tfrac{\bar\nu_{z_i}^{k}-\nu_{z_{i-1}}^{k}}
             {(\bar u_{s_i}^{k}-\nu_{z_{i-1}}^{k})^{2}}\Bigr]
\Bigr]
\boldsymbol{K}_{u_{s_i}\nu_{z_i}}^{jk}
\\
&\quad+
\rho_{i-1}^{j}(\bar u_{s_i}^{j}-\nu_{z_{i-1}}^{j})
\Bigl(\bar\nu_{z_i}^{k}-\nu_{z_{i-1}}^{k}
      +\tfrac{P_{i-1}^{k}}{\rho_{i-1}^{k}}
       \tfrac{1}{\bar u_{s_i}^{k}-\nu_{z_{i-1}}^{k}}\Bigr)
\boldsymbol{K}_{\nu_{z_i}\nu_{z_i}}^{jk}
\end{aligned}
\end{equation}

\noindent
\textit{Density–shock velocity cross-covariance}\\
\noindent Using the derivatives in Eqs.~\eqref{drho_du}–\eqref{d2rho_du_dv} one obtains
\begin{equation}
\label{K_rho_us}
\boldsymbol{K}_{\rho_iu_{s_i}}^{jk} =
\frac{\rho_{i-1}^{j}\bigl(\bar\nu_{z_i}^{j}-\nu_{z_{i-1}}^{j}\bigr)}
     {\bigl(\bar u_{s_i}^{j}-\bar\nu_{z_i}^{j}\bigr)^{2}}
\boldsymbol{K}_{u_{s_i}u_{s_i}}^{jk}
\;+\;
\frac{\rho_{i-1}^{j}\bigl(\bar u_{s_i}^{j}-\nu_{z_{i-1}}^{j}\bigr)}
     {\bigl(\bar u_{s_i}^{j}-\bar\nu_{z_i}^{j}\bigr)^{2}}
\boldsymbol{K}_{u_{s_i}\nu_{z_i}}^{jk}
\end{equation}

\noindent
\textit{Density–particle velocity cross-covariance}\\
\noindent Using the same derivatives gives
\begin{equation}
\label{K_rho_vz}
\boldsymbol{K}_{\rho_i\nu_{z_i}}^{jk} =
\frac{\rho_{i-1}^{j}\bigl(\bar\nu_{z_i}^{j}-\nu_{z_{i-1}}^{j}\bigr)}
     {\bigl(\bar u_{s_i}^{j}-\bar\nu_{z_i}^{j}\bigr)^{2}}
\boldsymbol{K}_{u_{s_i}\nu_{z_i}}^{jk}
\;+\;
\frac{\rho_{i-1}^{j}\bigl(\bar u_{s_i}^{j}-\nu_{z_{i-1}}^{j}\bigr)}
     {\bigl(\bar u_{s_i}^{j}-\bar\nu_{z_i}^{j}\bigr)^{2}}
\boldsymbol{K}_{\nu_{z_i}\nu_{z_i}}^{jk}
\end{equation}

\noindent
\textit{Density–energy cross-covariance}\\
\noindent Using the derivatives in Eqs.~\eqref{drho_du}–\eqref{d2rho_du_dv} and \eqref{dE_du}–\eqref{d2E_du_dv} one obtains
\begin{equation}
\label{K_rho_E}
\begin{aligned}
\boldsymbol{K}_{\rho_iE_i}^{jk}
&=
\frac{\rho_{i-1}^{j}\bigl(\bar\nu_{z_i}^{j}-\nu_{z_{i-1}}^{j}\bigr)}
     {\bigl(\bar u_{s_i}^{j}-\bar\nu_{z_i}^{j}\bigr)^{2}}
\Bigl[-\frac{P_{i-1}^{k}}{\rho_{i-1}^{k}}
       \frac{\bar\nu_{z_i}^{k}-\nu_{z_{i-1}}^{k}}
            {(\bar u_{s_i}^{k}-\nu_{z_{i-1}}^{k})^{2}}\Bigr]
\boldsymbol{K}_{u_{s_i}u_{s_i}}^{jk}
\\[4pt]
&\quad+
\Bigl[
\frac{\rho_{i-1}^{j}(\bar\nu_{z_i}^{j}-\nu_{z_{i-1}}^{j})}
     {(\bar u_{s_i}^{j}-\bar\nu_{z_i}^{j})^{2}}
\Bigl(\bar\nu_{z_i}^{k}-\nu_{z_{i-1}}^{k}
      +\tfrac{P_{i-1}^{k}}{\rho_{i-1}^{k}}
       \tfrac{1}{\bar u_{s_i}^{k}-\nu_{z_{i-1}}^{k}}\Bigr)
\\[-2pt]
&\hspace{18mm}+
\frac{\rho_{i-1}^{j}(\bar u_{s_i}^{j}-\nu_{z_{i-1}}^{j})}
     {(\bar u_{s_i}^{j}-\bar\nu_{z_i}^{j})^{2}}
\Bigl[-\tfrac{P_{i-1}^{k}}{\rho_{i-1}^{k}}
       \tfrac{\bar\nu_{z_i}^{k}-\nu_{z_{i-1}}^{k}}
             {(\bar u_{s_i}^{k}-\nu_{z_{i-1}}^{k})^{2}}\Bigr]
\Bigr]
\boldsymbol{K}_{u_{s_i}\nu_{z_i}}^{jk}
\\[4pt]
&\quad+
\frac{\rho_{i-1}^{j}(\bar u_{s_i}^{j}-\nu_{z_{i-1}}^{j})}
     {(\bar u_{s_i}^{j}-\bar\nu_{z_i}^{j})^{2}}
\Bigl(\bar\nu_{z_i}^{k}-\nu_{z_{i-1}}^{k}
      +\tfrac{P_{i-1}^{k}}{\rho_{i-1}^{k}}
       \tfrac{1}{\bar u_{s_i}^{k}-\nu_{z_{i-1}}^{k}}\Bigr)
\boldsymbol{K}_{\nu_{z_i}\nu_{z_i}}^{jk}
\end{aligned}
\end{equation}

\textit
\textit{Energy–shock velocity cross-covariance}\\
\noindent Using the derivatives in Eqs.~\eqref{dE_du}–\eqref{d2E_du_dv} one obtains  
\begin{equation}
\label{K_Eus}
\begin{aligned}
\boldsymbol{K}_{E_iu_{s_i}}^{jk}
&=
\Bigl[-\frac{P_{i-1}^{j}}{\rho_{i-1}^{j}}
       \frac{\bar\nu_{z_i}^{j}-\nu_{z_{i-1}}^{j}}
            {(\bar u_{s_i}^{j}-\nu_{z_{i-1}}^{j})^{2}}\Bigr]
\boldsymbol{K}_{u_{s_i}u_{s_i}}^{jk}
\\[4pt]
&\quad+
\Bigl[\bar\nu_{z_i}^{j}-\nu_{z_{i-1}}^{j}
      +\frac{P_{i-1}^{j}}{\rho_{i-1}^{j}}
       \frac{1}{\bar u_{s_i}^{j}-\nu_{z_{i-1}}^{j}}\Bigr]
\boldsymbol{K}_{\nu_{z_i}u_{s_i}}^{jk}
\end{aligned}
\end{equation}

\noindent
\textit{Energy–particle velocity cross-covariance}\\
\noindent Using the same derivatives gives  
\begin{equation}
\label{K_Evz}
\begin{aligned}
\boldsymbol{K}_{E_i\nu_{z_i}}^{jk}
&=
\Bigl[-\frac{P_{i-1}^{j}}{\rho_{i-1}^{j}}
       \frac{\bar\nu_{z_i}^{j}-\nu_{z_{i-1}}^{j}}
            {(\bar u_{s_i}^{j}-\nu_{z_{i-1}}^{j})^{2}}\Bigr]
\boldsymbol{K}_{u_{s_i}\nu_{z_i}}^{jk}
\\[4pt]
&\quad+
\Bigl[\bar\nu_{z_i}^{j}-\nu_{z_{i-1}}^{j}
      +\frac{P_{i-1}^{j}}{\rho_{i-1}^{j}}
       \frac{1}{\bar u_{s_i}^{j}-\nu_{z_{i-1}}^{j}}\Bigr]
\boldsymbol{K}_{\nu_{z_i}\nu_{z_i}}^{jk}
\end{aligned}
\end{equation}

\section{Additional Thermodynamic Constraints}
\label{app:Thermodynamic_Cosntraints}

Importantly, the material as represented by its state variables must satisfy the laws of thermodynamics. These are distinguished into two conditions. The first constraint is the thermodynamic consistency condition, which establishes a relationship between the thermodynamic variables through a single variable known as the Helmholtz free energy, $F$, defined as
\begin{equation}\label{Helmholtz}
F = E - T S
\end{equation}
where \(S\) denotes the entropy. The first law of thermodynamics states that changes in the free energy are given in the differential form \(dF = -S dT - PdV\). Thus, the pressure and energy can be related to the Helmholtz free energy by
\begin{equation}
    P = -\left.\frac{\partial F}{\partial V}\right|_T, \quad
    E = F - T \left.\frac{\partial F}{\partial T}\right|_V
\end{equation}
Taking the derivatives \(\partial P/\partial T\) and \(\partial E/\partial V\) yields the thermodynamic consistency constraint
\begin{equation}
P = T \frac{\partial P}{\partial T} - \frac{\partial E}{\partial V}
\label{consistency_constraint}
\end{equation}

The second thermodynamic constraint, also known as thermodynamic stability, is obtained from the second law of thermodynamics, which requires convexity of the Helmholtz free energy with respect to the thermodynamic variables. As a result, the specific heat (\(c_V\)) and isothermal compressibility (\(\kappa_T\)) are positive quantities. This is consistent with the expectation that energy must be added to the system to raise its temperature and that an increase in pressure decreases the volume. In this regard, the thermodynamic stability constraints are given as
\begin{equation}\label{stability_1}
\left(\frac{\partial^2 S}{\partial E^2}\right)_V = -\frac{1}{T^2} \left(\frac{\partial T}{\partial E}\right)_V 
= \frac{1}{T^2 c_V} \leq 0 
\quad \Longleftrightarrow \quad
\left(\frac{\partial T}{\partial E}\right)_V \geq 0
\end{equation}
\begin{equation}\label{stability_2}
\left(\frac{\partial^2 F}{\partial V^2}\right)_T = -\left(\frac{\partial P}{\partial V}\right)_T = \frac{1}{V \kappa_T} \geq 0 
\quad \Longleftrightarrow \quad
\left(\frac{\partial P}{\partial V}\right)_T \leq 0
\end{equation}

Clearly, a GP formulation that satisfies the generalized Hugoniot jump conditions in Eqs.~\eqref{eqn:gen_hugoniot_density}, \eqref{Hugoniot_pressure_gen} and \eqref{Hugoniot_energy_gen} should also be consistent with these thermodynamic constraints. In this regard, from the maximal set of all possible hyperparameters, $\boldsymbol{\theta}$, in Eq.~\eqref{gp_kernel}, a thermodynamically consistent subset is identified by enforcing the equality constraint in Eq.~\eqref{consistency_constraint} and the inequalities in Eqs.~\eqref{stability_1} and \eqref{stability_2}.\textbf{}

\section{Molecular Dynamics Simulations for Shockwave Propagation in SiC}
\label{sec:MD_simulations}

This section describes the Molecular Dynamics simulations performed to generate and process the data for training the proposed GP model for SiC.

SiC has multiple polytypes, and we chose 3C-SiC (also called $\beta$-SiC) due to its rich shock response. It has a cubic zinc-blende structure with a lattice constant of 4.354 \text{Å}. Each silicon atom is tetrahedrally bonded to four carbon atoms, and vice versa, forming a strong covalent network. MD shock simulations were conducted using the open source code LAMMPS\cite{lammps} from Sandia National Laboratories with an interatomic potential developed by Vashishta et al.~\cite{potential}, which has been very successful in capturing basic physical and mechanical properties of SiC.

\begin{figure}[!ht]
    \centering
    \includegraphics[width=0.5\linewidth]{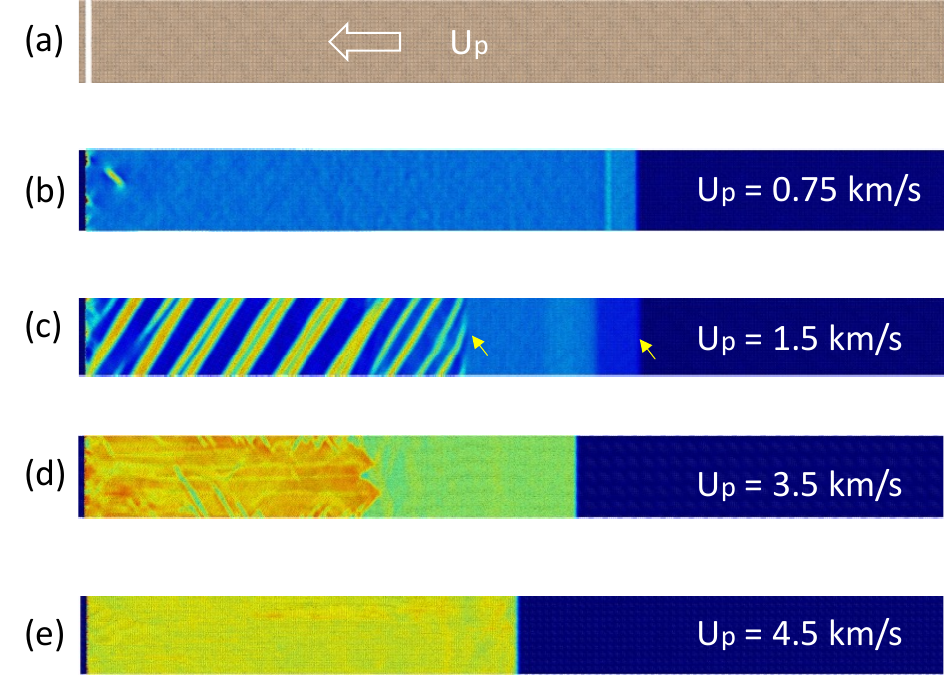}
    \caption{Shock wave simulations in MD: (a) Schematic setup; 
(b) Elastic wave propagation from $u_p=0.75$ km/s. (c) Dual elastic-plastic wave formation from $u_p=1.5$ km/s. (d) Dual plastic-phase transformation wave formation from $u_p=3.5$ km/s. (e) Overdriven shock propagation. In all figures, colors show atomic shear strains.}
    \label{fig:system}
\end{figure}

The shock simulations used the reverse ballistic approach, where a pre-equilibrated target sample is launched into a fixed piston with a velocity $-u_p$. The impact launches a shock wave in the target in the direction opposite to $u_p$. The temporal evolution of the system is described via adiabatic MD (NVE ensemble) with a timestep of 1 fs. The (001) plane was chosen as the impact plane, thus the shock wave propagates in the [001] direction. The target is formed by replicating the 8-atom unit cell 60$\times$60$\times$1020 times along [001], [010], and [001], respectively, resulting in cross-sectional dimensions of 25nm $\times$ 25nm and a length of $\sim$450 nm. The fixed piston has the same cross section as the target and a thickness of 2 nm. The initial gap between the piston and target is 2 nm. The total system has 28,587,552 atoms and periodic boundary conditions are imposed in the transverse directions. A schematic setup is shown in Figure \ref{fig:system}(a), together with snapshots of shocks of different strengths.

\subsection{Temperature, Density, and Stress Profiles}
We studied a series of particle velocities, ranging from 0.25 km/s to 6.0 km/s in increments of 0.25 km/s. The simulations were stopped at 36 ps, before the shock wave reached the free surface of the target. For weak shocks, we observe a purely elastic wave, but increasing the shock strength beyond the HEL results in plastic deformation and a phase transition. The snapshots colored by shear strain shown in Figure~\ref{fig:system} clearly display the wave fronts.

The shock front is marked by sharp changes in local properties, including density, velocity, temperature, and stress. To analyze the MD simulations and extract consistent and reproducible information to train the GP model, we developed and implemented an automatic workflow consisting of several steps described here and in the following sub-section. The first step is to divide the simulation domain into thin bins and average properties or parameters over the bins. We performed a one-dimensional binning in the shock direction with a bin size of 2 nm. The temperature, density, and stress profiles are then determined as follows.

The temperature per bin is calculated using the relationship between kinetic energy and temperature, i.e., 
\begin{equation}
    \Sigma_i m_iv_i^2 = k_B T N_{dof}
\end{equation}
where the sum runs over the atoms $i$ within the bin, $m_i$ and $v_i$ are, respectively, the atomic mass and velocity referenced to the center of mass velocity of the bin, $k_B$ is the Boltzmann constant, and $N_{dof}=3N-6$ is the total number of degrees of freedom for the $N$ atoms in a bin. We averaged the per-bin temperature over 1 ps and output the temperature data for each bin accompanying with the spatial bin center. Similarly, we can obtain stress (or pressure) profiles. The atomic stress tensor ($\sigma_{\alpha\beta}$) can be calculated by using the following stress definition\cite{equation_1, equation_2}:
\begin{equation}
    \sigma^{i}_{\alpha\beta} = -\frac{1}{\Omega^i}[m^iv^i_{\alpha}v^i_\beta + \frac{1}{2}\Sigma_{j\neq i}(r^{ij}_{\alpha}F^{ij}_{\beta}) + \frac{1}{3}\Sigma_{j,k}(r^{ij}_{\alpha}F^{ijk}_{\beta})
\end{equation}
where $\Omega$, $m$, and $v$ are the atomic volume, mass, and velocity subtracted by the bin center of mass motion, respectively.  The first term is the kinetic energy contribution. The second term is the pairwise energy contribution, and the third term is the contribution from three-body interactions. $r$ and $F$ represent the atom position and the corresponding force on the central atom and those involved in the specific interaction. The stresses were averaged over the number of atoms per bin and further averaged over a period of 1 ps and then output for each bin together with the spatial bin center. The density is in units of g/cm$^3$ and is calculated as the total mass of atoms in a bin divided by the volume of the bin.

\begin{figure}[!ht]
    \centering
    \includegraphics[width=1.0\linewidth]{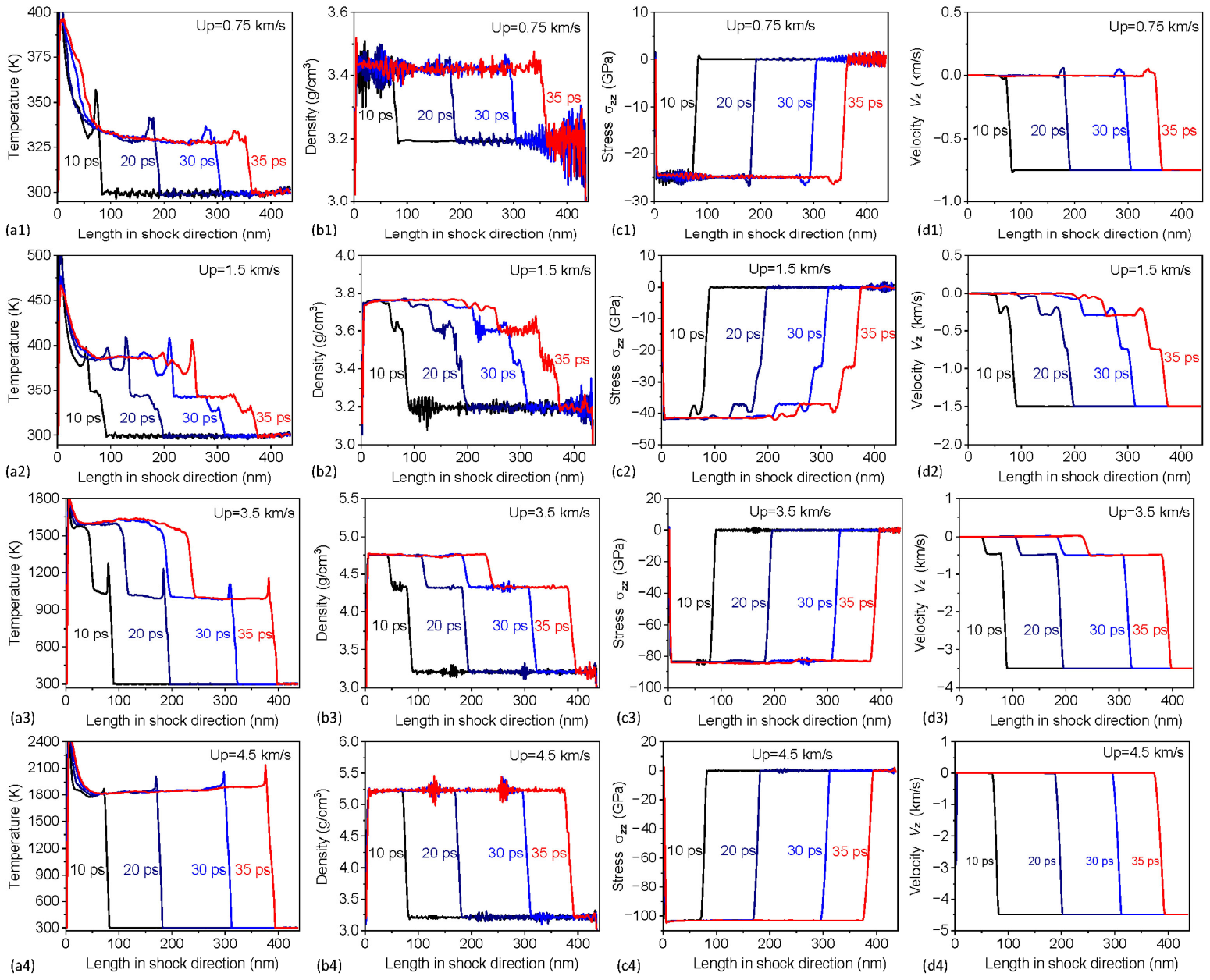}
    \caption{Profiles of (a) temperature, (b) mass density, (c) total stress $\sigma_{zz}$ along shock direction (z: [001]), and (d) particle velocity from MD shock simulations at (1) $u_p=0.75$ km/s, (2) $u_p=1.5$ km/s, (3) $u_p=3.5$ km/s, and (4) $u_p=4.5$ km/s.}
    \label{fig:profiles}
\end{figure}

Figure \ref{fig:profiles} shows temperature, density, stress, and velocity profiles for selected particle velocities: $u_p=$ 0.75 km/s, 1.5 km/s, 3.5 km/s, and 4.5 km/s. The wave propagation can be easily identified from the sharp change in properties. As detailed below, the shock velocity $u_s$ is obtained from the shock front position at various times. For $u_p=0.75$ km/s in Figure~\ref{fig:profiles}(a1), (b1), and (c1), a single elastic shock front is observed having relatively constant density and stress behind the wave front. Increasing the shock strength to 1.5 km/s in Figure~\ref{fig:profiles}(a2), (b2), and (c2) results in two wave fronts. The trailing plastic wave has higher density, temperature, and stress than the leading elastic wave. The shear strain snapshot in Figure~\ref{fig:system}(c) clearly shows the corresponding plastic strain pattern. For $u_p=3.5$ km/s, we observe two waves from the temperature and density profiles in Figures~\ref{fig:profiles}(a3), (b3). Interestingly, the second wave is barely noticeable in the stress profile in Figure~\ref{fig:profiles}(c3). This is because energy is mostly converted to internal heat without increasing pressure when SiC undergoes a phase transformation, as previously reported~\cite{previous-work-1, previous-work-2}. For particle velocity $u_p=4.5$ km/s, there is clearly only one phase transformation wave from each of the profiles.

\subsection{Shock Property Extraction}
Extracting meaningful data from shock simulations, including the identification of wave fronts and the calculation of wave speeds and averaged properties of the shock material, is typically done by hand. This results in tedious manual work and poor reproducibility. To address these issues, we automated and scripted not only the data extraction from the MD simulations but also their subsequent analysis. Property profiles, such as those shown in Figure \ref{fig:profiles}, are utilized to calculate shock speeds and the materials properties on both sides of the shockwave. Our approach involves identifying the shock front(s) and then using this information at various times to calculate the quantities of interest. Importantly, we use the jump conditions to verify the calculations. The Hugoniot jump conditions apply to steadily propagating waves with properties as shown in Figure \ref{schematic_shock_2}.

\begin{figure}[!ht]
    \centering
    \includegraphics[width=0.8\linewidth]{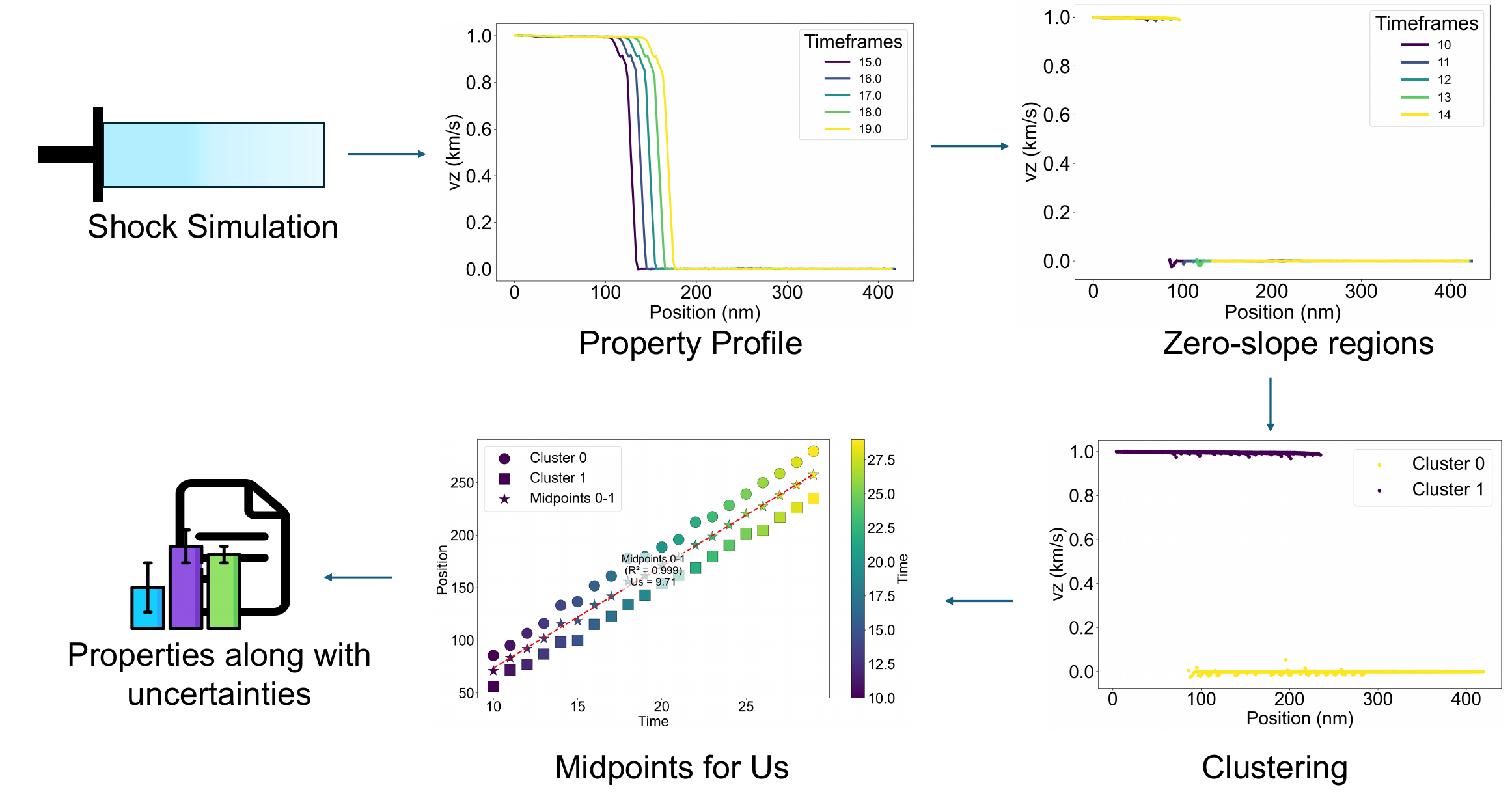}
    \caption{The workflow we utilize to get shock speeds from a property profile obtained from LAMMPS simulation. Then we cluster the zero-sloped regions toghether. Using the cluster we get the quasi-equilibrium state properties. Midpoints between different cluster boundaries are computed and shock speed is obtained as shown. At the end we save each property along with the standard deviation in the cluster averaging.}
    \label{fig:workflow}
\end{figure}

The automated workflow illustrated in Figure~\ref{fig:workflow} takes the property profiles as inputs and returns the shock speed(s) and the shock state properties for a given $u_p$. To identify the shock front(s) from the profiles, we first identify regions where properties are constant by performing a cluster analysis of the property values; these clusters denote the various states of the system (e.g., unshocked, elastically deformed, plastically deformed). 
While multiple properties can be used to identify shock fronts, not all property profiles are equally noise-proof for automated clustering. We studied various properties with different clustering algorithms and parameters at different piston velocities to determine which property would be most suitable for clustering. We find that particle velocity and density are the best properties to identify the number of clusters and hence the wavefronts. We used the hierarchical density-based spatial clustering of applications with noise (HDBSCAN)\cite{hdbscan} algorithm due to its known applications in noise-containing data implemented within the scikit-learn library\cite{scikit}. 

\begin{figure}[!ht]
    \centering
    \includegraphics[width=\linewidth]{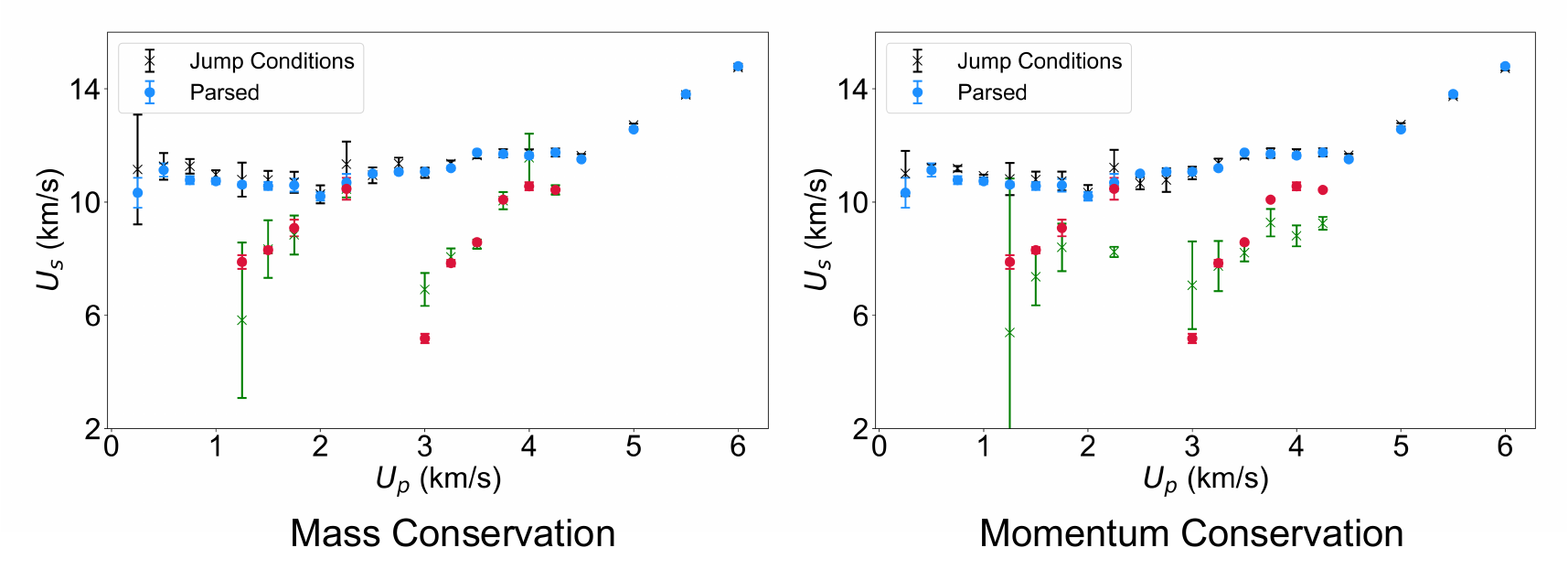}
    \caption{Comparing parsed $u_s$ with the values calculated from the Rankine-Hugoniot jump conditions. The error bars for the jump conditions are obtained using error propagation.}
    \label{fig:jump_conditions}
\end{figure}

Using the resulting clusters at a given time, we define shock fronts as the midpoint between the start and end of subsequent regions. The change in position of these midpoints with time is the shock speed U$_s$, we obtain average velocities by fitting the position vs time data with linear functions. The $R^2$ obtained from the linear fitting of the midpoints was used to determine the optimal property for clustering. To validate the automatic data analysis and confirm the steady nature of the shocks, we calculate $u_s$ from the averaged pre- and post-shock properties using the jump conditions and compare it with the results of the linear fit. If the difference is larger than the propagated error, the process stops and gives a warning. Figure \ref{fig:jump_conditions} shows that, apart from a few points, the data match the jump conditions well. Higher errors in momentum conservation arise from error propagation in the stresses, which have the highest deviations.

\begin{figure}[!ht]
    \centering
    \includegraphics[width=0.65\linewidth]{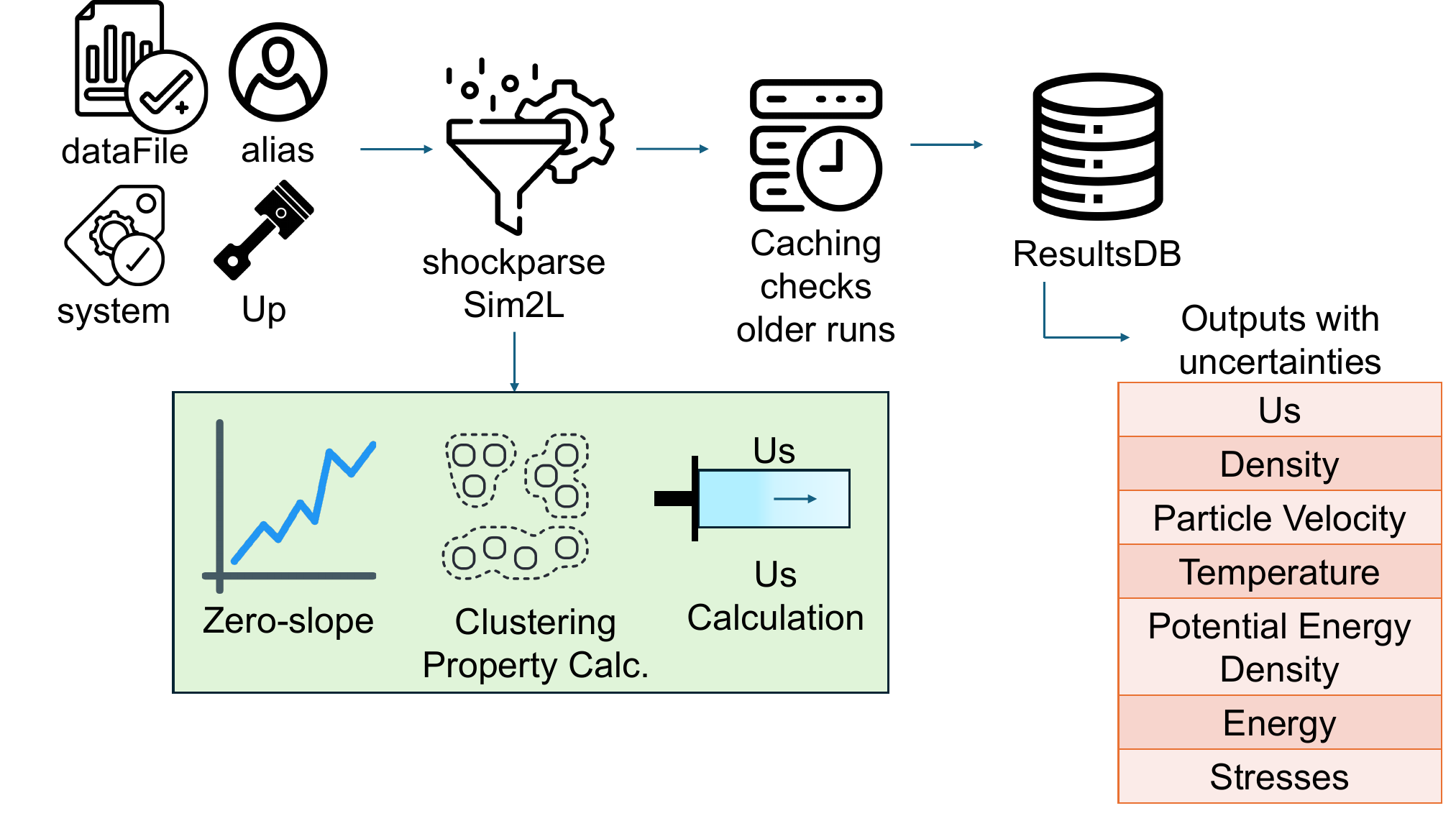}
    \caption{Workflow for the online published tool that saves the results in a universally indexed database for future querying.}
    \label{fig:tool-workflow}
\end{figure}

The workflow generates quasi-equilibrium state values for properties: stress, velocity, density, temperature, and energy, including their standard deviations and shock speeds. The workflow has been published as an open-source tool on nanoHUB,\cite{hunt2022sim2ls} accessible for online simulations\cite{shockparse-tool}, where researchers can analyze their own data. The tool takes the trajectory file and piston velocity as input, along with metadata information such as alias, system, and type of simulation (e.g. “Momentum Mirror”), so that it can be easily retrieved later. It calculates all the properties mentioned above and saves them with the standard deviation in clustering. Everything is stored in a universally indexed database, which also implements caching to avoid data redundancy. This aligns with our goal of making workflows Findable, Accessible, Interoperable, and Reusable (FAIR) for the advancement of the scientific community. The inputs, outputs, and workflow of the tool are highlighted in Figure \ref{fig:tool-workflow}.

\subsection{Up-Us Datasets}
\begin{figure}[!ht]
    \centering
    \includegraphics[width=0.5\linewidth]{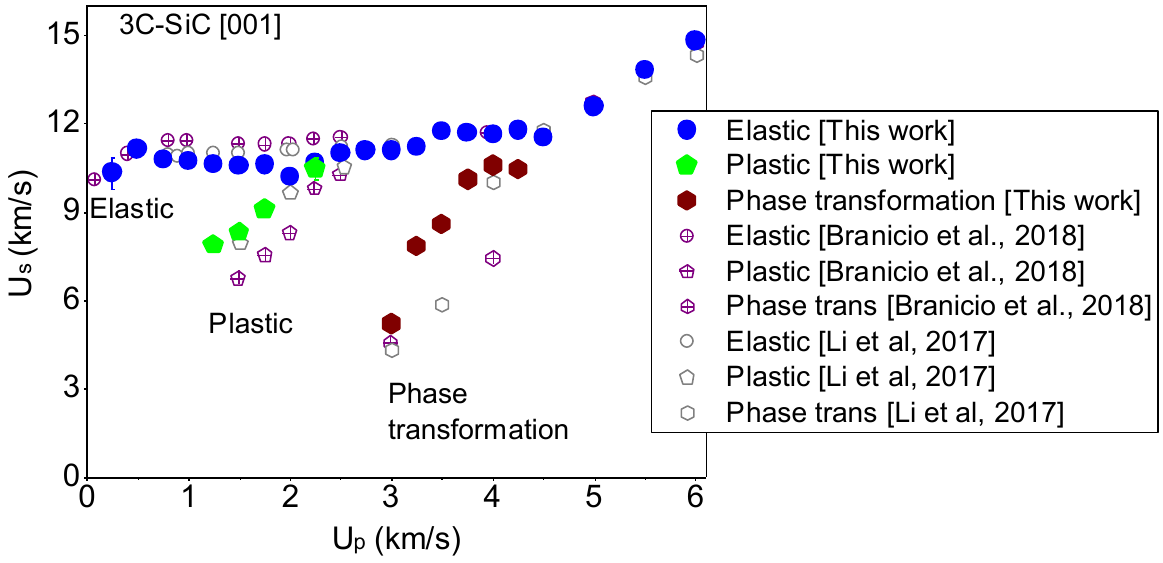}
    \caption{Shock velocity versus particle velocity from this work compared with previous results.}
    \label{fig:comparison}
\end{figure}

The final $u_s$ vs. $u_p$ data set is shown in Figure~\ref{Us_vs_Up} and compared with data available from the literature in Figure~\ref{fig:comparison}. 
The data shows good agreement with past studies, again highlighting the regimes of interest comprised of the three branches: elastic wave, plastic wave, and phase transformation wave. The plastic wave begins to show at particle velocity around $u_p=1.25$ km/s. Its propagation speed increases with increasing particle velocity until the particle velocity reaches $u_p=2.5$ km/s. The phase transformation wave starts when the particle velocity is slightly higher than $u_p=2.5$ km/s. Its propagation speed also increases with increasing particle velocity until the particle velocity reaches $u_p=4.25$ km/s. When the particle velocity is above $u_p=4.5$ km/s, there is a single overdriven wave, whose propagation speed shows a linear increase with particle velocity.

\bibliographystyle{elsarticle-num}
\biboptions{sort&compress}
\bibliography{HEL_GP}
\end{document}